\documentclass[letterpaper,english,aps,prl,reprint,amsmath,amssymb]{revtex4-2}
\usepackage[T1]{fontenc}
\usepackage[latin9]{inputenc}
\usepackage{color}
\usepackage{babel}
\usepackage{amsmath}
\usepackage{amssymb}
\usepackage{graphicx}
\usepackage{microtype}
\usepackage[unicode=true,pdfusetitle,
 bookmarks=true,bookmarksnumbered=true,bookmarksopen=true,bookmarksopenlevel=1,
 breaklinks=false,pdfborder={0 0 0},pdfborderstyle={},backref=false,colorlinks=true]
 {hyperref}

\makeatletter

\pdfpageheight\paperheight
\pdfpagewidth\paperwidth

\providecommand{\tabularnewline}{\\}

\usepackage{braket}
\usepackage{mathtools}
\usepackage{graphicx}
\hypersetup{linkcolor = blue,
            urlcolor  = blue,
            citecolor = blue,
            anchorcolor = blue}

\makeatother

\begin{document}
\title{Non-monotonic temperature dependence and first-order phase transition of relaxation times in molecular spin}
\author{Le Tuan Anh Ho}
\email{chmhlta@nus.edu.sg}

\affiliation{Department of Chemistry, National University of Singapore, 3 Science Drive 3 Singapore 117543}
\author{Liviu Ungur}
\email{chmlu@nus.edu.sg}

\affiliation{Department of Chemistry, National University of Singapore, 3 Science Drive 3 Singapore 117543}
\author{Liviu F. Chibotaru}
\email{liviu.chibotaru@kuleuven.be }

\affiliation{Theory of Nanomaterials Group, Katholieke Universiteit Leuven, Celestijnenlaan 200F, B-3001 Leuven, Belgium}
\date{\today}
\begin{abstract}
We derive a simple system of equations to describe the magnetization relaxation of a molecular spin in weak interaction with a thermal bath for the whole temperature domain. Using this for the intermediate temperature domain where the transition from coherent to incoherent relaxation occurs, we find that the slowest relaxation mode shows a first-order phase transition. Associated with this transition, an unusual non-monotonic temperature-dependence of the relaxation rate of this mode is also demonstrated. Contrary to the popular belief, this non-monotony gives rise to a peculiar but observable behavior where increasing temperature will not only result in a smaller rate of the slowest relaxation mode but also may lead to a slower decaying of the magnetization after some relaxing time. Additionally, it is also shown that magnetization relaxation in this intermediate temperature domain can only be accurately described by a bi- or tri-exponential form. The physical reason underlying these features can be attributed to the role of the quantum tunneling effect and different but comparative relaxation modes. A simple experiment to confirm our findings on the first-order phase transition and the non-monotony of the relaxation rate is accordingly proposed.
\end{abstract}
\maketitle
\global\long\def\hmt{\mathcal{H}}%
\global\long\def\vt#1{\mathbf{#1}}%

\global\long\def\chip{\chi'}%
\global\long\def\chipp{\chi''}%

\global\long\def\ketbra{\ket{m}\bra{m}}%

Relaxation of a spin system in a thermal bath is a fundamental problem in physics and material science. With recent advances in molecular magnetism toward the use of magnetic molecules in high density magnetic storage, quantum information processing, spintronic devices, and molecular-based multifunctional magnetic materials \citep{Leuenberger2001,Hill2003,Vincent2012,Bogani2008,Gatteschi2003,Sessoli1993,Tejada2001,Wernsdorfer2002,Godfrin2017,Moreno-Pineda2018,Wernsdorfer2019,Goodwin2017,Guo2018,sanvito2011molecular,abellan2015hybrid,Chen2022a,ohkoshi201490,coronado2005pressure,Ashebr2022,Serrano2020,sato1996photoinduced,maspoch2003nanoporous,train2008strong,coronado2006synthesis,kurmoo2007superconducting,Moreno-Pineda2021}, a thorough understanding of the effect of the thermal bath on the spin system becomes meaningful than ever. This understanding, undoubtedly, will facilitate the realization of a new generation of molecular spin-based materials in quantum technology and nanotechnology.

Making use of Born and Markov approximations, dynamics of a spin system in a thermal (phonon) bath can be approached via the density matrix formalism with Redfield master equation. This equation has been ubiquitously used in molecular magnetism to deal with monodomain magnetic particles, especially those systems exhibiting quantum tunneling of magnetization (QTM) \citep{Garanin1997,Leuenberger2000,Gatteschi2006,Garanin2011,lunghi2017role,Ho2018,gu2020origins}. Solutions of this equation not only provides relaxation and decoherence rate induced by the spin-phonon interaction but also elucidate the role of the quantum tunneling process. At high temperature limit, practical use of the equation allows a quantification of the incoherent quantum tunneling of magnetization and incoherent relaxation accordingly \citep{Garanin1997,Leuenberger2000,Gatteschi2003,Gatteschi2006}. Meanwhile, at low temperature limit, this equation converges back to the canonical description of the time-evolution of a two-level system where ground states population shows a Rabi oscillation due to the effect of the coherence transition within the ground (quasi-) doublet \citep{griffiths2005introduction,Garanin2011}. For the intermediate temperature domain where the transition from incoherent to coherent relaxation occurs, as far as we are aware, a theoretical investigation is still lacking. Hence, in this Letter, we focus on the relaxation behavior of a multilevel spin system $S\left(J\right)$ in interaction with a thermal bath in the mentioned intermediate temperature domain where there exists a transition from incoherent to coherent relaxation. Our description for the relaxation in molecular spin here is also valid for the whole temperature range.

\emph{Model} -- A multilevel spin system in a ligand field with the following generic spin Hamiltonian is treated:
\begin{multline}
\hmt=\sum_{m^{\mathrm{th}}}\left(\varepsilon_{m}+\frac{W_{m}}{2}\right)\ket{m}\bra{m}+\left(\varepsilon_{m}-\frac{W_{m}}{2}\right)\ket{m'}\bra{m'}\\
+\sum_{m^{\mathrm{th}}}\left(\frac{\Delta_{m}}{2}\ket{m}\bra{m'}+\frac{\Delta_{m}^{*}}{2}\ket{m'}\bra{m}\right)+\sum_{n^{\mathrm{th}}}\varepsilon_{n}\ket{n}\bra{n},
\end{multline}
where $m^{\mathrm{th}}$, $n^{\mathrm{th}}$ denote doublets and singlets respectively, $\Delta_{m}$ is the tunnel splitting of the corresponding doublet regardless its intrinsic (non-Kramers system) or field-induced (Kramers system) nature. Choosing $z$-axis along the main magnetic axis of the molecular spin, $W_{m}$ is then the energy bias between localized states $\ket{m}$ and $\ket{m'}$ due to the effect of the longitudinal component of the magnetic field \citep{Ho2022a}.  It is also supposed that the ground states are of doublet type and the magnetic field is not so strong to switch the energy order of the doublets/singlets.

Relaxation of the spin is governed by the Redfield master equation in its non-secular form \citep{Garanin2011,Blum1996}. However, since the relaxation rate is typically much slower than the energy gaps between doublets/singlets, we can further apply the semi-secular approximation \citep{Garanin2011,Ho2017} to obtain:
\begin{align}
\frac{d\rho_{mm}}{dt} & =-\frac{i}{2}\left(\Delta_{m}\rho_{m'm}-\Delta_{m}^{*}\rho_{mm'}\right)\nonumber \\
 & \qquad+\sum_{k}R_{mm,kk}\rho_{kk}+\sum_{k}R_{mm,kk'}\rho_{kk'},\label{eq:semi-secular-Redfield-equation1}\\
\frac{d\rho_{mm'}}{dt} & =-i\left[W_{m}\rho_{mm'}+\frac{\Delta_{m}}{2}\left(\rho_{m'm'}-\rho_{mm}\right)\right]\nonumber \\
 & \qquad+\sum_{k}R_{mm',kk}\rho_{kk}+\sum_{k}R_{mm',kk'}\rho_{kk'}.\label{eq:semi-secular-Redfield-equation2}
\end{align}
where $R_{mn,m'n'}$ are matrix elements of the Redfield super-operator,  $\rho$ is the reduced density matrix of the spin system. These equations are applicable for both doublets and singlets with a convention that there is no corresponding state $\ket{n'}$ or $\Delta_{n}$ for the singlet $\ket{n}$.

During relaxation process, lifetime of excited states is relatively short comparing to the total relaxation time \citep{Garanin1997,Leuenberger2000,Gatteschi2006}. This is either due to fast dynamics of the populations between states at high temperature or large emission rates to lower energy states at low temperature. This allows us to consider the stationary limit for all \emph{excited} doublets/singlets $m^{\mathrm{th}}$:
\begin{equation}
\frac{d\rho_{mm}}{dt}=\frac{d\rho_{m'm'}}{dt}=0,\,\frac{d\rho_{mm'}}{dt}=\frac{d\rho_{m'm}}{dt}=0.\label{eq:stationary limit}
\end{equation}
The symmetry between localized states $\ket{m}$ and $\ket{m'}$ as the ``up'' and ``down'' state of $m^{\mathrm{th}}$ will give us the following approximated solutions without \emph{explicitly} solving Eq. \eqref{eq:stationary limit} \citep{SupplementalMaterial}:
\begin{align}
\rho_{mm} & =A_{mm}^{11}\rho_{11}+A_{mm}^{1'1'}\rho_{1'1'}+A_{mm}^{11'i}\rho_{11'i},\\
\rho_{m'm'} & =A_{mm}^{11}\rho_{1'1'}+A_{mm}^{1'1'}\rho_{11}-A_{mm}^{11'i}\rho_{11'i},\\
\rho_{mm'r} & =A_{mm'}^{11'r}\rho_{11'r},\\
\rho_{mm'i} & =A_{mm'}^{11i}\left(\rho_{11}-\rho_{1'1'}\right)+A_{mm'}^{11'i}\rho_{11'i},
\end{align}
where $\rho_{mm'r}$ and $\rho_{mm'i}$ are the real and imaginary part of $\rho_{mm'}$ and all coefficients $A_{mn}^{ij}$ are real-valued parameters. Substituting these into Eqs. (\ref{eq:semi-secular-Redfield-equation1}-\ref{eq:semi-secular-Redfield-equation2}) for the ground doublets \citep{SupplementalMaterial}, we obtain the key system of equations for relaxation of a spin system in a thermal bath: 
\begin{gather}
\frac{dX_{1}}{dt}=-\Gamma_{e}X_{1}-2\left(\Delta_{1r}\rho_{11'i}-\Delta_{1i}\rho_{11'r}\right),\label{eq:dX1}\\
\frac{d\rho_{11'r}}{dt}=-\gamma_{11'}\rho_{11'r}+W_{1}\rho_{11'i}-\frac{\Delta_{1i}}{2}X_{1},\label{eq:drho11'r}\\
\frac{d\rho_{11'i}}{dt}=-W_{1}\rho_{11'r}-\gamma_{11'}\rho_{11'i}+\frac{\Delta_{1r}}{2}X_{1},\label{eq:drho11'i}
\end{gather}
where $X_{1}\equiv\rho_{11}-\rho_{1'1'}$ is the ground doublet population difference; $\gamma_{11'}=-R_{11',11'}$ is the escape rate of the ground doublet population; $\Delta_{1r}$ ($\Delta_{1i}$) is the real (imaginary) part of $\Delta_{1}$. Meanwhile, $\Gamma_{e}\equiv2\Gamma_{11'}+2\sum_{n^{\mathrm{th}}\ne1^{\mathrm{st}}}\left[\left(\Gamma_{1,n'}A_{nn}^{11}+\Gamma_{1,n}A_{nn}^{1'1'}\right)+2R_{11,nn'}^{i}A_{nn'}^{11i}\right]$ plays the role as the relaxation rate at zero tunnel splitting and results from two contributions: direct process rate $2\Gamma_{11'}$, and the effective relaxation rate via excited doublets/singlets (the remainder term of $\Gamma_{e}$). Here we denote $\Gamma_{mn}\equiv R_{mm,nn}$ as the transition rate from $\ket{n}$ to $\ket{m}$.

Clearly, solution of Eqs. (\ref{eq:dX1}-\ref{eq:drho11'i}) can be found in the exponential form $\sum c_{i}e^{-\lambda_{i}t}$. As the corresponding characteristic polynomial is only of the third order, its roots can thus be written in the analytical form \citep{SupplementalMaterial}. Here we first analyze the resonance case, $W_{1}=0$. For non-resonance case $W_{1}\ne0$, since it is hardly to infer any physical implications from the lengthy analytical forms of the roots of the third order polynomial equation, we resort to analyzing its numerical solutions.

\emph{Spin relaxation at resonance }-- For $W_{1}=0$, Eqs. (\ref{eq:dX1}-\ref{eq:drho11'i}) give three relaxation modes with corresponding rates: 
\begin{align}
\lambda_{1,2} & =\Gamma_{e}+\gamma\mp\sqrt{\gamma^{2}-\Delta_{1}^{2}}\equiv\Gamma_{e}+\Gamma_{1,2}^{\mathrm{tn}},\,\lambda_{3}=\gamma_{11'}.\label{eq:resonance rates}
\end{align}
where $\Delta_{1}\equiv\sqrt{\Delta_{1r}^{2}+\Delta_{1i}^{2}}$  and $\gamma\equiv\left(\gamma_{11'}-\Gamma_{e}\right)/2$. The quantity $\gamma$ can also represent for the decoherence rate since it is half of the escape rate $\gamma_{11'}$ corrected by the effect of the canonical relaxation $\Gamma_{e}$. For convenience, we name it the decoherence rate hereinafter.  It should also be noted that from our companion work \citep{Ho2022c}, we found that QTM can be considered as a driven and damped quantum harmonic oscillator where the quantities $\Delta_{1}$, $\gamma$, $\zeta\equiv\gamma/\Delta_{1}$ play the role of the undamped angular frequency, viscous damping coefficient, and damping ratio of the tunneling oscillation in the ground doublet respectively. The components $\Gamma_{1,2}^{\mathrm{tn}}=\gamma\mp\sqrt{\gamma^{2}-\Delta_{1}^{2}}$ can thus be considered as resulting from the tunneling effect of the population in the ground doublet. Therefore, rate of each relaxation mode is simply an addition of the relaxation via canonical relaxation channels, $\Gamma_{e}$, and quantum tunneling one. 

\begin{figure}
\begin{centering}
\includegraphics[width=8.5cm]{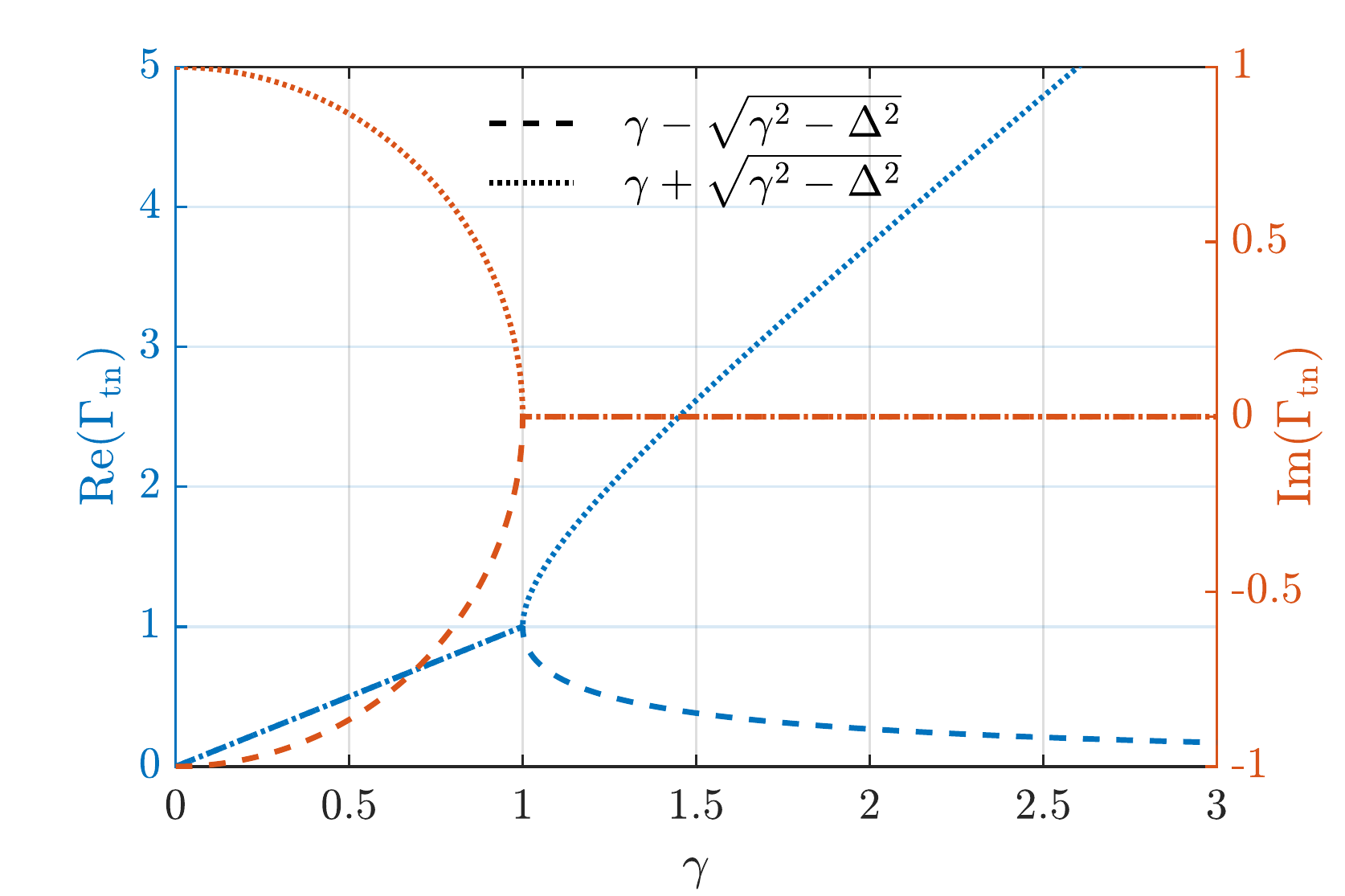}
\par\end{centering}
\caption{Tunneling-related rate $\Gamma_{1,2}^{\mathrm{tn}}\left(\gamma\right)$ at resonance in $\Delta_{1}=1$ unit. \label{fig:Tunneling_rate}}
\end{figure}

Using the typical initial condition for the density matrix element in the recovery magnetization measurement $\left(X_{1},\rho_{11'r},\rho_{11'i}\right)\mid_{t=0}=\left(1,0,0\right)$, we obtain \citep{SupplementalMaterial}: 
\begin{align}
M\left(t\right) & \approx\frac{M_{0}e^{-\Gamma_{e}t}}{2\sqrt{\gamma^{2}-\Delta_{1}^{2}}}\left(\Gamma_{2}^{\mathrm{tn}}e^{-\Gamma_{1}^{\mathrm{tn}}t}-\Gamma_{1}^{\mathrm{tn}}e^{-\Gamma_{2}^{\mathrm{tn}}t}\right),\label{eq:M(t) at resonance}
\end{align}

Interestingly, the magnetization relaxation at resonance is a combination of two relaxation modes $\lambda_{1,2}$. This is manifested clearly in the intermediate temperature domain when $\gamma\sim\mathcal{O}\left(\Delta_{1}\right)$ as can be seen from Fig. \ref{fig:Tunneling_rate} where the tunneling rate $\Gamma_{1,2}^{\mathrm{tn}}$ is plotted. Within this intermediate domain, not only relaxation rates $\lambda_{1,2}$ of two relaxation modes but also their contributions to $M\left(t\right)$, which are intriguingly approximately inversely proportional to the tunneling-related components $\Gamma_{1,2}^{\mathrm{tn}}$, are of the same order of magnitude. We can conclude that the \emph{bi-exponential} form, Eq. \eqref{eq:M(t) at resonance}, must be respected to describe correctly the magnetization relaxation in this temperature domain.

 Magnetization relaxation behavior at some limits can also be worked out from Eq. \eqref{eq:resonance rates} and \eqref{eq:M(t) at resonance} as well. Indeed, the magnetization relaxation of the spin is reduced to one dominant mode at high temperature due to $\lambda_{1}\ll\lambda_{2}$ where it relaxes incoherently with a total rate $\lambda_{1}=\Gamma_{e}+\gamma-\sqrt{\gamma^{2}-\Delta_{1}^{2}}\approx\Gamma_{e}+\Delta_{1}^{2}/2\gamma$ because of the large ground doublet escape rate $\gamma_{11'}$ and accordingly $\gamma$. Meanwhile, at low temperature, the magnetization relaxes coherently with a decay rate of $\left(\Gamma_{e}+\gamma\right)$ and a frequency $\sqrt{\Delta_{1}^{2}-\gamma^{2}}$, which asymptotically approaches the well-known Rabi oscillation $M\left(t\right)=M_{0}\cos\Delta_{1}t$ at very low temperature due to the phasing out of phonon-induced rate $\Gamma_{e}$ and $\gamma$.

As is clear from Fig. \ref{fig:Tunneling_rate} and Eq. \eqref{eq:resonance rates}, derivative of the tunneling rate relaxation rates $d\Gamma_{1,2}^{\mathrm{tn}}/d\gamma=\left(\sqrt{\gamma^{2}-\Delta_{1}^{2}}\mp\gamma\right)/\sqrt{\gamma^{2}-\Delta_{1}^{2}}$ shows a discontinuity at $\gamma=\Delta_{1}$. That is to say, the tunneling-related components $\Gamma_{1,2}^{\mathrm{tn}}$ and accordingly two slowest spin system relaxation rates $\lambda_{1,2}=\Gamma_{e}+\Gamma_{1,2}^{\mathrm{tn}}$ undergo a first-order phase transition at $\gamma=\Delta_{1}$ which corresponds to some transition temperature $T_{0}$ \footnote{Dependence of the transition temperature $T_{0}$ on the magnetic anisotropy of the spin system will be addressed in another separate publication.}. Since at this transition point, oscillating component in $M\left(t\right)$ begins to appear (disappear), we can thus physically associate this point as one separate the coherence ``phase'' and incoherence ``phase'' in the relaxation of the molecular spin system. It is also evident that the arising of this first-order phase transition is caused by the existence of multiple quantum tunneling modes since $\Gamma_{e}$ is the same in three rates $\lambda_{i}$. 

\begin{figure}
\begin{centering}
\includegraphics[width=8.5cm]{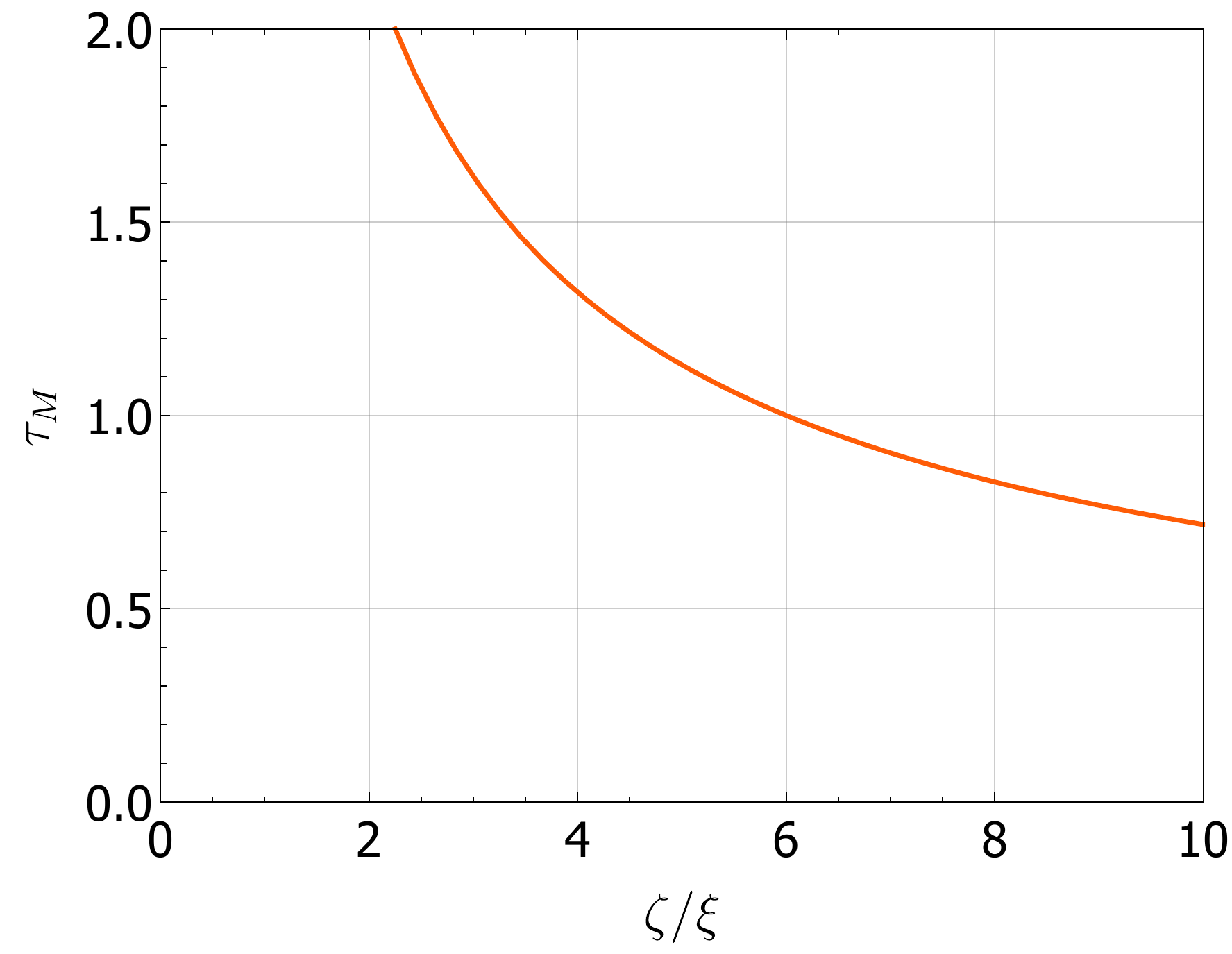}
\par\end{centering}
\caption{The time $\tau_{M}\left(\zeta/\xi\right)$ (in $\Delta_{1}^{-1}$ unit) from which a slower relaxation of $M\left(t,T_{0}+\delta T\right)$ comparing to $M\left(t,T_{0}\right)$ is observed in zero external magnetic field. \label{fig:tau_m}}
\end{figure}

Interestingly, in the positive proximity of this transition point,  the slowest relaxation rate $\lambda_{1}$ shows an unusual non-monotonic behavior where its value decreases with respect to a temperature increase. To prove this, let find the variation of $\lambda_{1}$ when the temperature makes a small change $\delta T$ from the transition temperature $T_{0}.$ Supposing that $\delta\gamma\left(T_{0}\right)=\zeta\delta T$ and $\delta\Gamma_{e}\left(T_{0}\right)=\xi\delta T$ where $\zeta,\xi>0$ we have: 
\begin{gather}
\delta\lambda_{1}\left(T_{0}\right)=\left[1-\sqrt{\frac{2\Delta_{1}\zeta}{\left(\zeta+\xi\right)^{2}\delta T}}\right]\left(\xi+\zeta\right)\delta T.
\end{gather}
Obviously, the inequality $\delta\lambda_{1}<0$ is always satisfied as long as $T$ is close to $T_{0}$ so that $0<\delta T<2\Delta_{1}\zeta/\left(\zeta+\xi\right)^{2}$. In other words, an increase in the temperature from $T_{0}$ will certainly lead to a decrease in the rate of the slowest relaxation mode. This unusual behavior of the slowest relaxation mode near the transition temperature $T_{0}$ may result in a peculiar behavior in the relaxation of the magnetization $M\left(t\right)$ where increasing the temperature will lead to a slower relaxation after some time $\tau_{M}=\Delta_{1}^{-1}\left(3+\sqrt{9+12\zeta/\xi}\right)\xi/2\zeta$. Fig. \ref{fig:tau_m} shows this dependence of $\tau_{M}$ as a function of $r=\zeta/\xi$. For $\tau_{M}$ in the observable time (smaller than $\tau_{\mathrm{relax}}\approx1/\Delta_{1}$), the condition $r\ge6$ must be met. In other words, the rate $\gamma$ needs to be much more sensitive to temperature than $\Gamma_{e}$ in the proximity of the transition point. 

It should also be noted that the first-order phase transition and non-monotony occurs at $\gamma=\Delta_{1}$ is likely not a coincidence. As mentioned, $\zeta=\gamma/\Delta_{1}$ plays the role of the viscous damping coefficient in the driven damped harmonic oscillator QTM \citep{Ho2022c}. In analogy with a damped harmonic oscillator, $\gamma=\Delta_{1}$ should corresponds to the critically damped case where population difference induced by the quantum tunneling effect return to its equilibrium as fast as possible without oscillating. Thus, any variation from this transition point will lead to a slower relaxation to the equilibrium. This is the physical reason behind the discovered non-monotony.

\emph{Spin relaxation out of resonance -- }From Eqs. (\ref{eq:dX1}-\ref{eq:drho11'i}), three relaxation rates $\lambda_{i}=\Gamma_{e}+\Gamma_{i}$, $i=1,2,3$, can be straightforwardly obtained (cf. Supplemental Material \citep{SupplementalMaterial} for their explicit form). The corresponding time-dependent magnetization is:

\begin{multline}
M\left(t\right)\approx M_{0}e^{-\Gamma_{e}t}\left[\frac{\Gamma_{2}\Gamma_{3}-\Delta_{1}^{2}}{\left(\Gamma_{1}-\Gamma_{2}\right)\left(\Gamma_{1}-\Gamma_{3}\right)}e^{-\Gamma_{1}t}\right.\\
+\frac{\Gamma_{1}\Gamma_{3}-\Delta_{1}^{2}}{\left(\Gamma_{2}-\Gamma_{1}\right)\left(\Gamma_{2}-\Gamma_{3}\right)}e^{-\Gamma_{2}t}\\
\left.+\frac{\Gamma_{1}\Gamma_{2}-\Delta_{1}^{2}}{\left(\Gamma_{3}-\Gamma_{1}\right)\left(\Gamma_{3}-\Gamma_{2}\right)}e^{-\Gamma_{3}t}\right].\label{eq:M(t)}
\end{multline}
As can be seen, the first part $M_{0}e^{-\Gamma_{e}t}$ describes the magnetization relaxation without the tunneling effect while the latter embraces this one. Similar to the resonance case, relaxation rates are simply summation of the effective relaxation rate via canonical relaxation channels, $\Gamma_{e}$, and three quantum tunneling rates, $\Gamma_{i}$, occurring within the ground doublet. At resonance, the above formula simply reduces to Eq. \eqref{eq:M(t) at resonance}. Meanwhile, at large magnetic field or high temperature, one relaxation mode is much slower than others and the magnetization relaxation is governed by only one relaxation mode with a familiar rate $\lambda_{1}\approx\Gamma_{e}+2\Delta_{1}^{2}\gamma/\left(W_{1}^{2}+4\gamma^{2}\right)$. In the case of small magnetic field, magnetization relaxation is similar to the resonance case with a small correction of the second order of magnitude $\mathcal{O}\left(W^{2}\right)$ \citep{Ho2022c}. 

\begin{figure}
\centering{}\includegraphics[width=8.5cm]{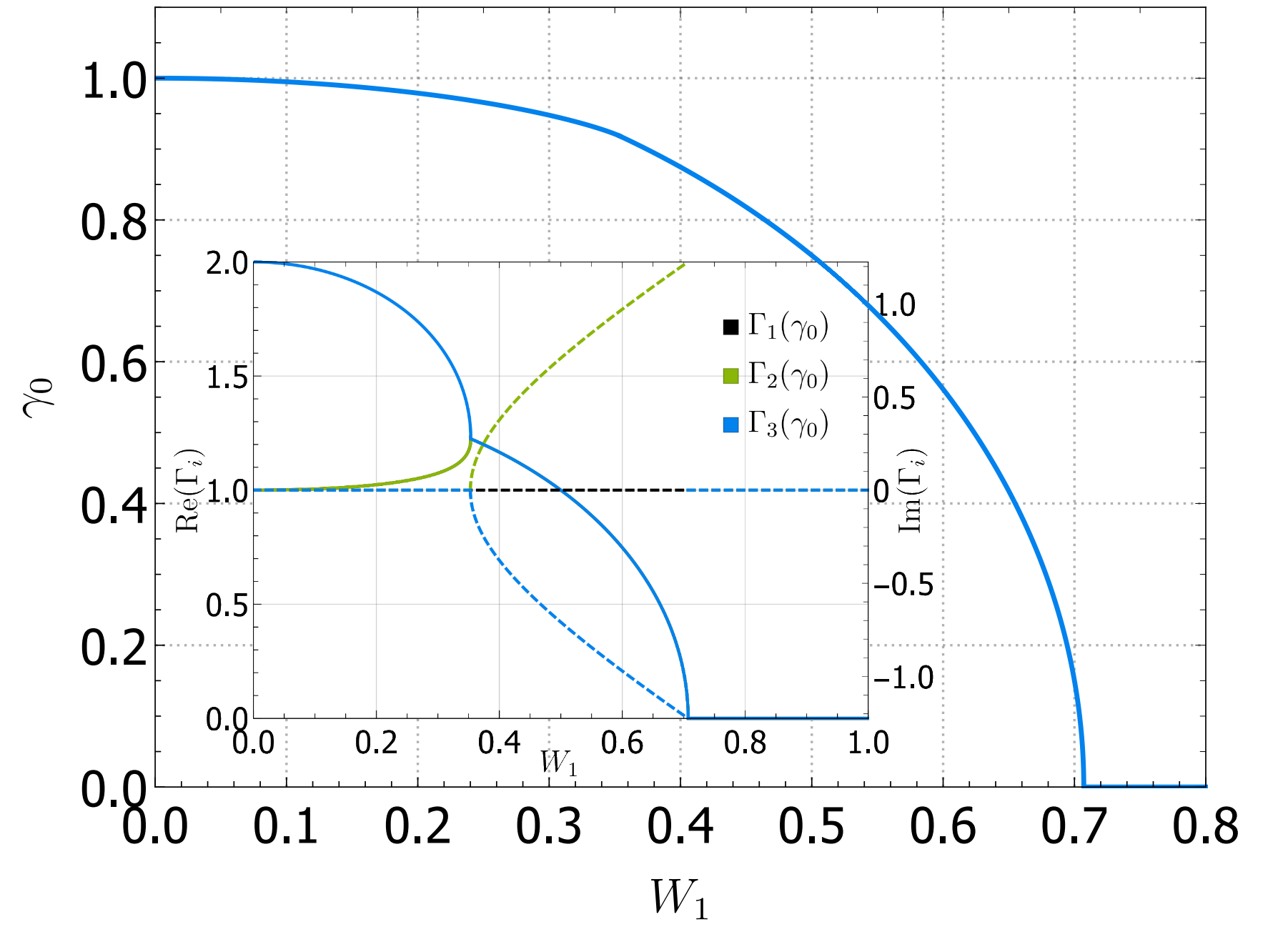}\caption{Transition decoherence rate $\gamma_{0}$ as a function of the energy bias $W_{1}$ in the ground doublet in $\Delta_{1}=1$ unit. Inset: Real (solid lines) and imaginary (dashed lines) component of tunneling rates $\Gamma_{i}$ corresponding to three relaxation modes at $\gamma_{0}$. \label{fig:Transition point}}
\end{figure}

Out of resonance, there still exists a specific $\gamma_{0}$ value (and accordingly a temperature $T_{0}$) where the slowest relaxation rate changes from complex to real value. At this point, at least two relaxation modes share the same decaying rate (see Fig. \ref{fig:Transition point-rates}). Hence, close to this transition point, using the tri-exponential form Eq. \eqref{eq:M(t)} is a must to describe correctly the magnetization relaxation. This transition point can also be considered as the place where the magnetization relaxation transits from coherent to incoherent manner due to the slowest relaxation mode change from purely decaying to decaying with oscillation and vice versa \citep{Ho2022c}. From our companion work \citep{Ho2022c}, we have: 
\begin{widetext}
\begin{eqnarray}
\gamma_{0}\left(W_{1}\right) & = & \begin{cases}
\Delta_{1}, & \text{if\,\,\,}W_{1}=0,\\
\frac{\left(\sqrt{\Delta_{1}^{2}-8W_{1}^{2}}+3\Delta_{1}\right)\sqrt{4W_{1}^{2}-\Delta_{1}\sqrt{\Delta_{1}^{2}-8W_{1}^{2}}+\Delta_{1}^{2}}}{8\sqrt{2}W_{1}}, & \text{if\,\,\,}0<W_{1}<\frac{\Delta_{1}}{2\sqrt{2}},\\
\frac{3}{2\sqrt{2}}\sqrt{\Delta_{1}^{2}-2W_{1}^{2}}, & \text{if\,\,\,}\frac{\Delta_{1}}{2\sqrt{2}}\le W_{1}<\frac{\Delta_{1}}{\sqrt{2}},\\
0, & \text{if\,\,\,}W_{1}\ge\frac{\Delta_{1}}{\sqrt{2}},
\end{cases}\label{eq:gamma0}
\end{eqnarray}
\end{widetext}

\noindent whose behavior is shown in Figure \ref{fig:Transition point}. As can be seen, the transition decoherence rate $\gamma_{0}$ rapidly decreases with the increase of the energy bias $W_{1}$ and disappears when $W_{1}\ge\Delta_{1}/\sqrt{2}$. Since tunneling splitting of the ground doublet is typically small, for a non-zero transition decoherence rate $\gamma_{0}$, and accordingly $T_{0}$, magnetic noise should be substantially reduced. From the physical meaning of the transition decoherence rate $\gamma_{0}$ mentioned above, our results here quantitatively states that either a high magnetic dilution or a sophisticated engineering to isolated the central spin from the internal field effect is required to maintain the coherence in the molecular spin. The corresponding quantum tunneling rates corresponding to three relaxation modes at the transition point is also shown in the inset of Fig. \ref{fig:Transition point}. It can be seen that when $W_{1}\le\Delta_{1}/2\sqrt{2}$, all rates are real and two slowest ones share the same value at the transition point. These three then gradually merge into one as $W_{1}$ approaches $\Delta_{1}/2\sqrt{2}$. Beyond this $W_{1}$ value, two of them are complex conjugates and all share the same decaying part .

\begin{figure}
\centering{}\includegraphics[width=8.5cm]{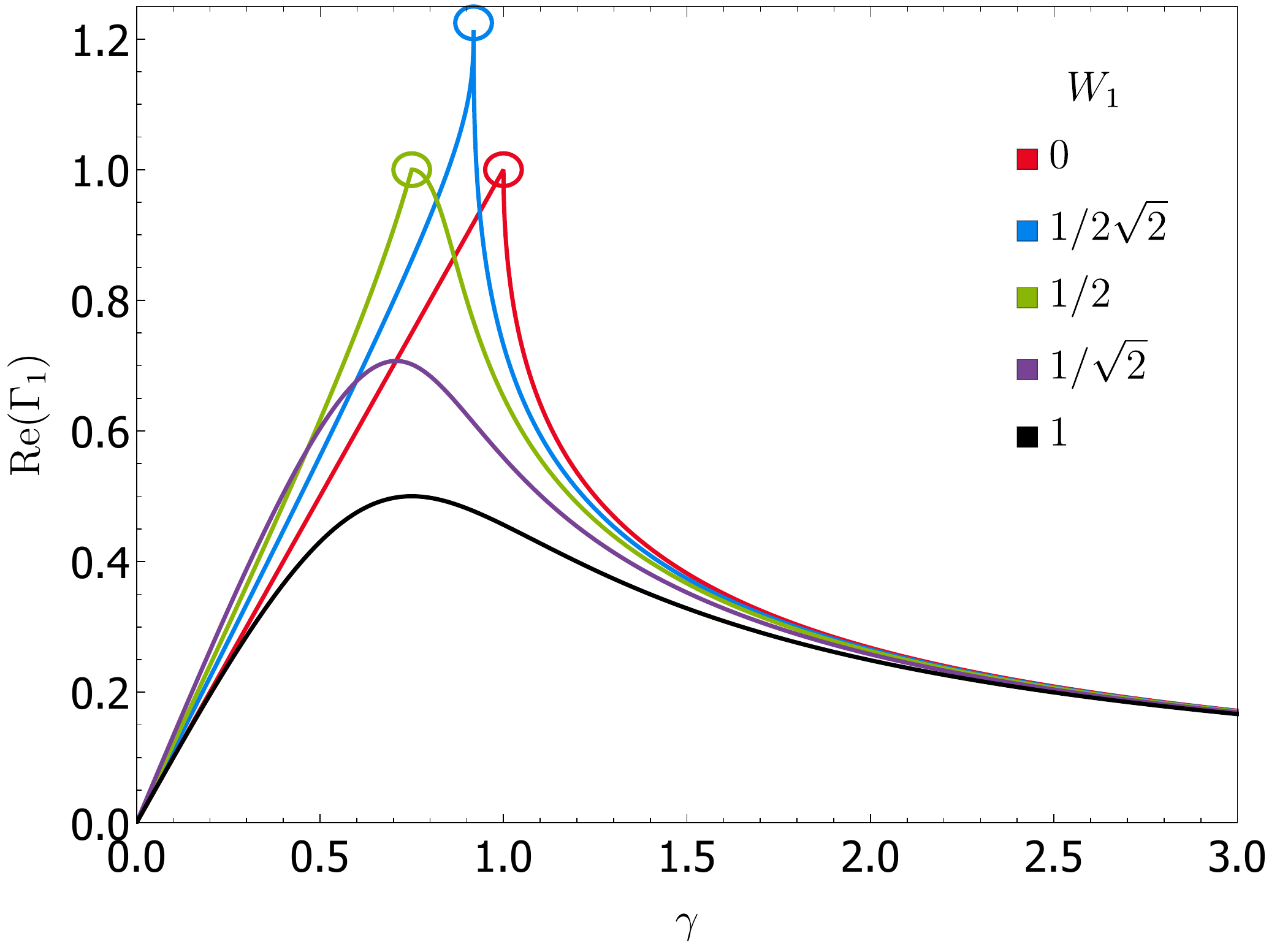}\caption{Real component of the tunneling rate corresponding to slowest relaxation mode as a function of the decoherence rate $\gamma$ in $\Delta_{1}=1$ unit. Colored circles indicates the transition point $\gamma_{0}$ and its corresponding slowest relaxation rate. \label{fig:Slowest relaxation mode}}
\end{figure}

Similar to resonance case, the slowest relaxation mode also undergoes a first-order phase transition at the transition point $\gamma_{0}$. This can be clearly seen in Fig. \ref{fig:Slowest relaxation mode} where the decaying part of tunneling rate of the slowest relaxation mode shows a discontinuity in its first derivative at non-zero $\gamma_{0}$ for $W_{1}<1/\sqrt{2}$ . This first-order phase transition corresponds to the transition from coherent to incoherent manner of the quantum tunneling rate which takes place at the transition point $\gamma_{0}$. It thus disappears when $\gamma_{0}=0$ as $W_{1}$ approaches $\Delta_{1}/\sqrt{2}$ as clearly seen. Since the total relaxation rate is the summation of the canonical one $\Gamma_{e}$ and the tunneling one $\Gamma_{\mathrm{tn}}$, the discontinuity in the first derivative of the total relaxation rate, or equivalently the first-order phase transition, is inherited from the tunneling one. 

Another interesting behavior of the transition from coherent to incoherent relaxation is the non-monotonic temperature dependence of the slowest total relaxation rate. Since the total relaxation rate of a mode equals to $\lambda=\Gamma_{e}+\Gamma_{\mathrm{tn}}$, we first summarize the quantum tunneling rate corresponding to the slowest relaxation mode $\Gamma_{\mathrm{tn}}^{\mathrm{sl}}$ for $\gamma\ge\gamma_{0}$ in the following:

\begin{align}
\Gamma_{\mathrm{tn}}^{\mathrm{sl}} & =\begin{cases}
\gamma-\sqrt{\gamma^{2}-\Delta^{2}}, & \text{if\,\,}W_{1}=0,\\
\frac{4\gamma}{3}+\frac{1+i\sqrt{3}}{6}\frac{3\Omega_{1}^{2}-4\gamma^{2}}{S}-\frac{1-i\sqrt{3}}{6}S, & \text{if\,\,}0<W_{1}\le\frac{\Delta_{1}}{2\sqrt{2}},\\
\frac{4\gamma}{3}-\frac{3\Omega_{1}^{2}-4\gamma^{2}}{3S}+\frac{1}{3}S, & \text{if\,\,}\frac{\Delta_{1}}{2\sqrt{2}}<W_{1},
\end{cases}\nonumber \\
\end{align}
where $\Omega_{1}=\sqrt{\Delta_{1}^{2}+W_{1}^{2}}$, $S\equiv\sqrt[3]{9\gamma\left(\Delta_{1}^{2}-2W_{1}^{2}\right)-8\gamma^{3}+3\sqrt{3}\sqrt{D}}$, and $D\equiv16\gamma^{4}W_{1}^{2}+\gamma^{2}\left(8W_{1}^{4}-20W_{1}^{2}\Delta_{1}^{2}-\Delta_{1}^{4}\right)+\Omega_{1}^{6}$.

\begin{figure}
\centering{}%
\begin{tabular}{c}
\includegraphics[width=8.5cm]{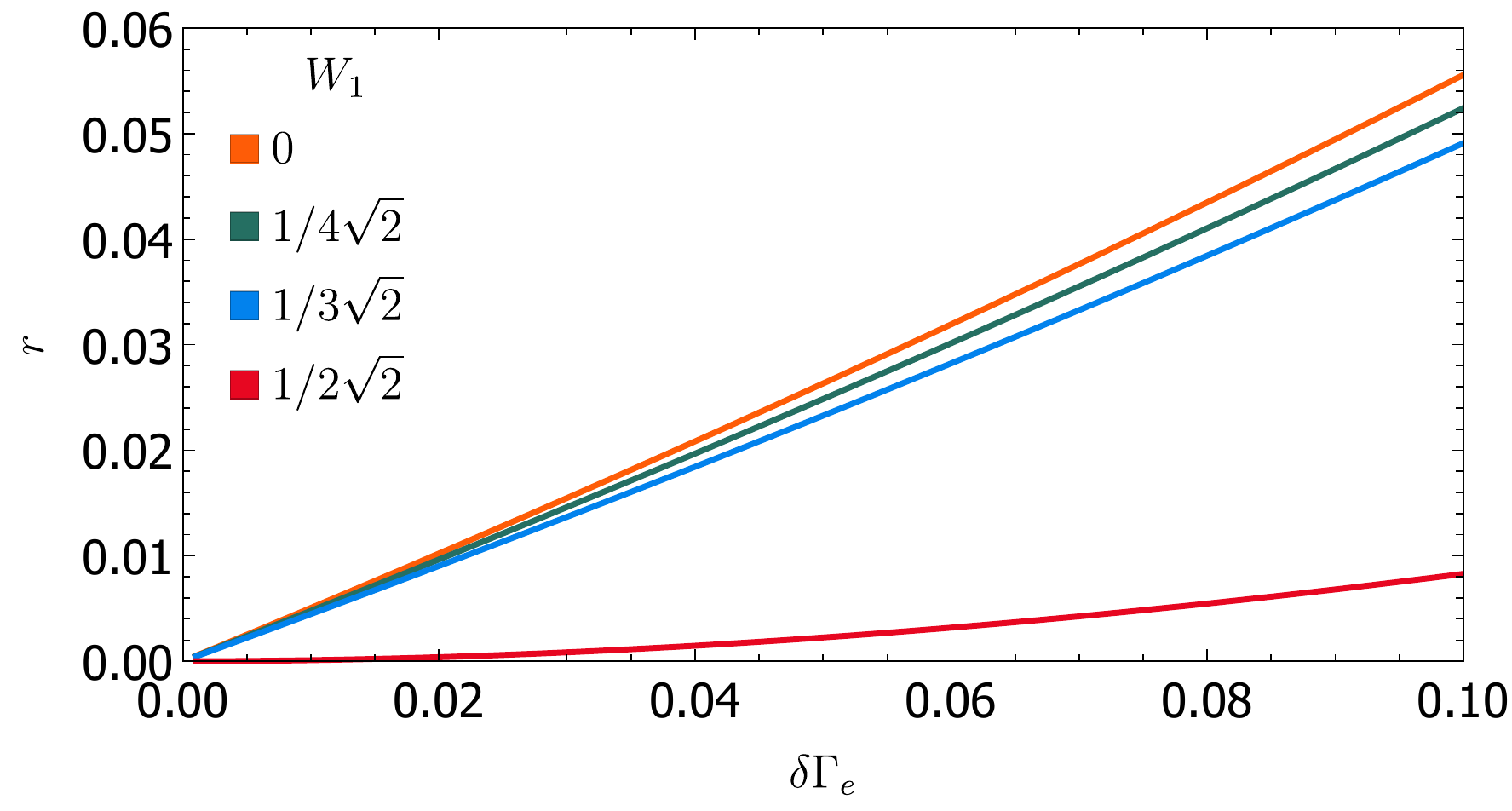}\tabularnewline
\includegraphics[width=8.5cm]{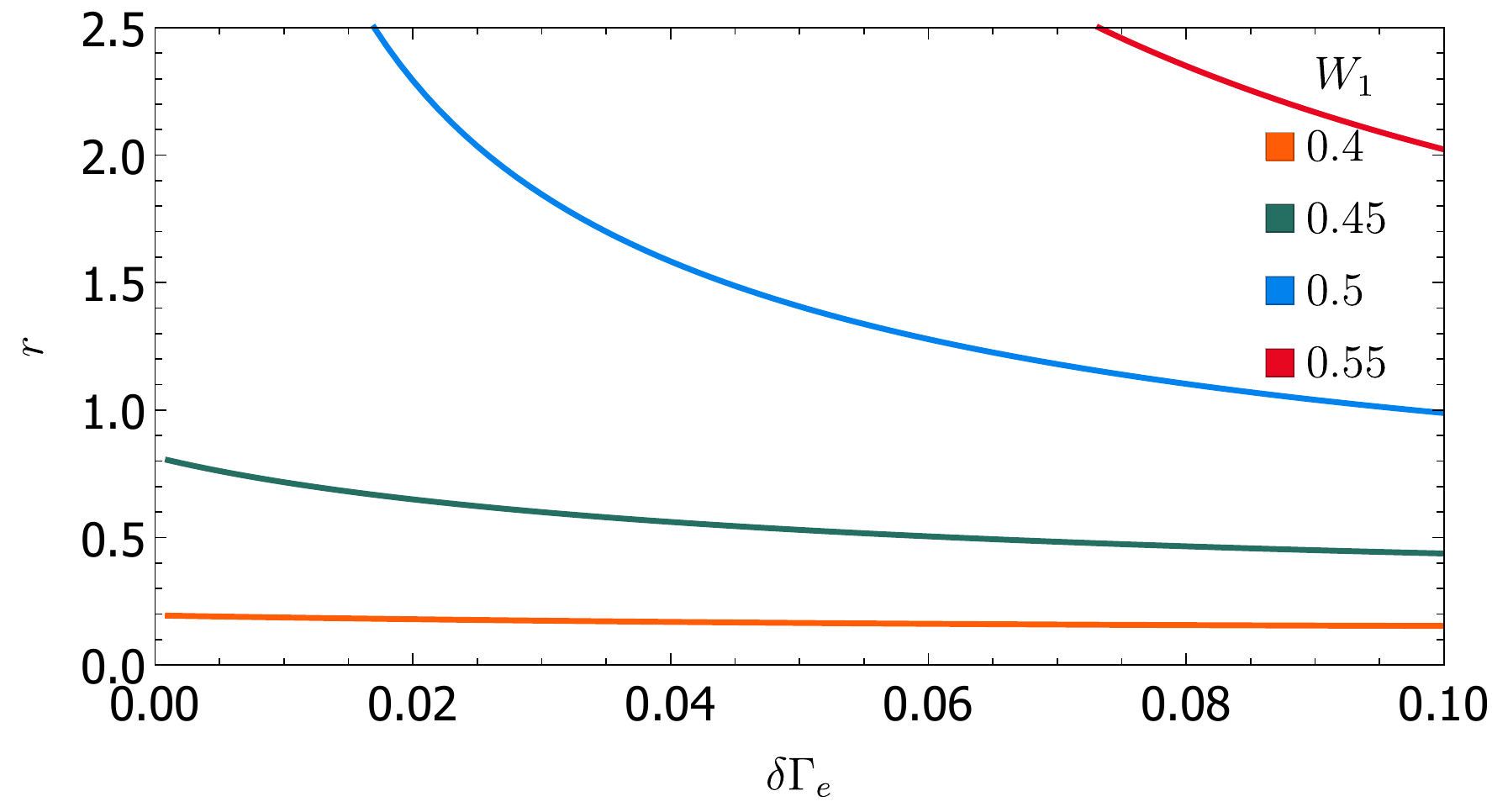}\tabularnewline
\end{tabular}\caption{Plot of the variation of the slowest relaxation mode $\delta\lambda_{1}=0$ at the transition point as a function of the $\delta\Gamma_{e}$ and the ratio $r=\zeta/\xi$ corresponding to different energy biases within the ground doublet in $\Delta_{1}=1$ unit. Here the area above each curve corresponds to the domain where $\delta\lambda_{1}\left(T_{0}\right)<0$. \label{fig:nonMonotony}}
\end{figure}

To demonstrate the non-monotonic temperature dependence of the slowest relaxation rate, i.e. an increasing temperature from the transition temperature results in a decrease in the total relaxation rate $\Gamma$ of the slowest relaxation mode, we consider a variation $\delta\lambda\equiv\lambda\left(\gamma_{0}+\delta\gamma\right)-\lambda\left(\gamma_{0}\right)$ when the decoherence rate $\gamma$ changes a small quantity $\delta\gamma>0$ from $\gamma_{0}$:
\begin{align}
\delta\lambda\left(\gamma_{0}\right) & =\delta\Gamma_{e}+\Gamma_{\mathrm{tn}}^{\mathrm{sl}}\left(\gamma_{0}+r\delta\Gamma_{e}\right)-\Gamma_{\mathrm{tn}}^{\mathrm{sl}}\left(\gamma_{0}\right).\label{eq:delta lambda - out of resonance}
\end{align}
where $r\equiv d\gamma/d\Gamma_{e}=\zeta/\xi$ defined previously. Since $\Gamma_{e}$ is the relaxation rate via canonical channels and thus is a monotonic function of the temperature, an increase in the temperature is equivalent to an increase of $\delta\Gamma_{e}$ and vice versa. Given a small (positive) value of $\delta\Gamma_{e}$, as long as a positive ratio $r$ can be found so that the inequality $\delta\lambda<0$ is satisfied then the non-monotonic temperature dependence of the slowest relaxation rate is proven. Equivalently, given a value of the ratio $r$, if we can find a domain of small $\delta\Gamma_{e}$ so that $\delta\lambda<0$ then this non-monotony is proven as well. In short, the proof now reduces to checking for the existence of a domain where $\delta\lambda<0$ in the first quadrant (top right) of the coordinate plane forming by $\left(\delta\Gamma_{e},r\right)$ given a minuscule $\delta\Gamma_{e}$. To fulfill this purpose, in principle we can analytically solve Eq. \ref{eq:delta lambda - out of resonance} in a similar way as the resonance case but for $\delta\Gamma_{e}$ instead of $\delta T$.  for example of the resonance case). However, due to its complexity, we resort to a numerical proof which is illustrated in Fig. \ref{fig:nonMonotony} where solutions of the equation $\delta\lambda_{1}=0$ corresponding to different values of the energy bias $W_{1}$ are shown. In this figure, the area above each curve corresponds to the domain of $\delta\lambda_{1}\left(\gamma_{0}\right)<0$. As can be seen, for all $r$ and $0\le W_{1}/\Delta_{1}\le1/2\sqrt{2}$, given $r$ we can always find a value of $\delta\Gamma_{e}$, and accordingly $\delta T$, so that $\delta\lambda_{1}<0$, i.e. an increase of temperature from the transition point will certainly results in a decrease in the slowest relaxation rate. In the case $1/2\sqrt{2}<W_{1}/\Delta_{1}\le1/\sqrt{2}$, the ratio $r$ must be larger than some specific value for $\delta\lambda_{1}/\delta T|_{T=T_{0}}<0$ and $r\rightarrow\infty$ as $W_{1}\rightarrow\Delta_{1}/\sqrt{2}$ and $\delta\Gamma_{e}\rightarrow0$. For $W_{1}>\Delta_{1}/\sqrt{2}$ where $\gamma_{0}=0$, no non-monotony at the transition point is found. From above and here, we can roughly state in short that when the tunneling effect is significant, the non-monotony and the first-order phase transition can be observed and vice versa.

Up to now, we have demonstrated that there may exist a first-order phase transition and a non-monotonic temperature dependence in the slowest relaxation mode in the intermediate temperature regime depending on the energy bias in the ground doublet. A simple experiment can be prepared to confirm these predictions. In this experiment, a strong diluted crystalline sample of molecular spins to reduce the effect of the magnetic noise is necessary. At each temperature, the sample  is magnetically saturated and its magnetization relaxation is measured. Temperature is reduced gradually until an oscillation in the magnetization relaxation is detected. Magnetization relaxation at those temperatures close to this special point is then fitted by a bi-exponential form (at resonance), or tri-exponential form (out-of-resonance) to separate the rates of different relaxation modes. The plots of these rates will reveal the non-monotony of the slowest relaxation mode as well as its first-order phase transition. Additionally, the tunneling splitting gap of the ground doublet can also be measured. Indeed, at resonance and this transition point, it is show that \citep{SupplementalMaterial,Ho2022c} the time-dependent magnetization relaxation is subject to the law $M\left(t\right)=M_{0}e^{-\Gamma t}\left[1+\Delta_{1}t\right]$. This law is then fitted with the magnetization relaxation data at the temperature where an oscillation in the magnetization relaxation is observed. The fitted parameter will then reveal the value of $\Delta_{1}$. 

In conclusion, the present work have contributed a simple system of equation and a sound framework to describe relaxation of a molecular spin in interaction with a thermal bath at any temperature. Applying this to the intermediate temperature domain where the transition from incoherent to coherent relaxation occurs, we discovered the first-order phase transition of the slowest relaxation rate of the magnetization relaxation at the transition point. Furthermore, at this point, increasing temperature will lead to an unusual non-monotonic temperature dependence of the slowest relaxation mode when the tunneling effect is significant, which may result in an observable behavior of magnetization relaxation after some period of time. More importantly, our work further pointed out that these special properties in the relaxation of the molecular spin system results from the quantum tunneling effect which behaves somewhat similar to a driven damped harmonic oscillator. In practice, we also shown that a bi- or tri-exponential form is required to describe accurately the magnetization relaxation in the intermediate temperature domain. This promises to have a potentially huge effect on the interpretation of the experimental relaxation data of molecular spins. Last but not least, these findings can be tested via our suggested simple experiment.

\begin{acknowledgments}
L. T. A. H. would like to thank Dr. Naoya Iwahara for helpful discussions. L. T. A. H. and L. U. acknowledge the financial support of the research projects R-143-000-A65-133, A-8000709-00-00 and A-8000017-00-00 of the National University of Singapore. Calculations were done on the ASPIRE-1 cluster (www.nscc.sg) under the projects 11001278 and 51000267. Computational resources of the HPC-NUS are gratefully acknowledged.
\end{acknowledgments}

\bibliographystyle{apsrev4-1}
\bibliography{references}

\end{document}


\title{Supplementary Material for \\
Non-monotonic temperature dependence and first-order phase transition of relaxation times in molecular spin}
\author{Le Tuan Anh Ho}
\email{chmhlta@nus.edu.sg}

\affiliation{Department of Chemistry, National University of Singapore, 3 Science Drive 3 Singapore 117543}
\author{Liviu Ungur}
\affiliation{Department of Chemistry, National University of Singapore, 3 Science Drive 3 Singapore 117543}
\author{Liviu F. Chibotaru}
\email{liviu.chibotaru@kuleuven.be }

\affiliation{Theory of Nanomaterials Group, Katholieke Universiteit Leuven, Celestijnenlaan 200F, B-3001 Leuven, Belgium}

\maketitle
\tableofcontents{}

\global\long\def\hmt{\mathcal{H}}%
\global\long\def\vt#1{\overrightarrow{#1}}%

\global\long\def\chip{\chi'}%
\global\long\def\chipp{\chi''}%

\global\long\def\tn{\mathrm{tn}}%

\section{Density matrix elements of the excited doublets/singlets in the stationary limit}

In the stationary limit of excited doublets/singlets $m^{\mathrm{th}}$, we have: 
\begin{equation}
\frac{d\rho_{mm}}{dt}=\frac{d\rho_{m'm'}}{dt}=0,\,\frac{d\rho_{mm'}}{dt}=\frac{d\rho_{m'm}}{dt}=0.\label{eq:stationary limit}
\end{equation}
If we consider the density matrix elements corresponding to the ground doublet as parameters, ones corresponding to excited levels can be found by solving this system of equations. Here all equations are formulated in localized basis \citep{Ho2022a}.

\subsection{Diagonal terms}

The symmetry between $\ket{m}$ and $\ket{m'}$ as ``up'' and ``down'' states corresponding to $m^{\mathrm{th}}$ doublet allows us to approximate the form of the diagonal density matrix elements for excited doublets/singlets of the investigated spin system as:
\begin{align}
\rho_{mm} & =A_{mm}^{11}\rho_{11}+A_{mm}^{1'1'}\rho_{1'1'}+A_{mm}^{11'}\rho_{11'}+A_{mm}^{1'1}\rho_{1'1},\label{eq:rhomm}\\
\rho_{m'm'} & =A_{mm}^{11}\rho_{1'1'}+A_{mm}^{1'1'}\rho_{11}+A_{mm}^{11'}\rho_{1'1}+A_{mm}^{1'1}\rho_{11'},\label{eq:rhom'm'}
\end{align}
where $A_{mn}^{ij}$ are parameters. These parameters can be found by numerically solving the system of equations \eqref{eq:stationary limit}. However, here we focus more on their properties instead of their value since these are not necessary for our work hereinafter.

In the stationary limit, we have $d\rho_{mm}/dt=d\rho_{m'm'}/dt=0$ for excited doublets/singlets. This also implies $d\left(\rho_{mm}+\rho_{m'm'}\right)/dt=0$ and $d\left(\rho_{11}+\rho_{1'1'}\right)/dt=0$, accordingly, 
\begin{gather}
\left(A_{mm}^{11'}+A_{mm}^{1'1}\right)\frac{d\rho_{11'r}}{dt}=0,
\end{gather}
at any moment $t$. This is satisfied if and only if $A_{mm}^{11'}+A_{mm}^{1'1}=0$. Since $\rho_{mm'}=\rho_{m'm}^{*}$ and $\rho_{mm}$, $\rho_{m'm'}$ are real, it is easy to see that $A_{mm}^{11'}=\left(A_{mm}^{1'1}\right)^{*}$. These two equations results in: 
\begin{equation}
\mathrm{Re}\left(A_{mm}^{11'r}\right)=0,
\end{equation}
thus we can rewrite Eq. \eqref{eq:rhomm} and \eqref{eq:rhom'm'} as: 
\begin{gather}
\rho_{mm}=A_{mm}^{11}\rho_{11}+A_{mm}^{1'1'}\rho_{1'1'}+A_{mm}^{11'i}\rho_{11'i},\label{eq:form_of_rho_mm}\\
\rho_{m'm'}=A_{mm}^{11}\rho_{1'1'}+A_{mm}^{1'1'}\rho_{11}-A_{mm}^{11'i}\rho_{11'i},\label{eq:form_of_rho_m'm'}
\end{gather}
where we have defined $A_{mm}^{11'i}\equiv-2\,\mathrm{Im}\left(A_{mm}^{11'}\right)$ and $\rho_{11'r}$, $\rho_{11'i}$ are respective real and imaginary part of $\rho_{11'}$.

\subsection{Off-diagonal terms}

Similarly,  off-diagonal density matrix elements for excited doublets/singlets can take the form: 
\begin{gather}
\rho_{mm'}=A_{mm'}^{11}\rho_{11}+A_{mm'}^{1'1'}\rho_{1'1'}+A_{mm'}^{11'}\rho_{11'}+A_{mm'}^{1'1}\rho_{1'1},\\
\rho_{m'm}=A_{mm'}^{11}\rho_{1'1'}+A_{mm'}^{1'1'}\rho_{11}+A_{mm'}^{11'}\rho_{1'1}+A_{mm'}^{1'1}\rho_{11'}.
\end{gather}
Since $\rho_{mm'}=\rho_{m'm}^{*}$, we have: 
\begin{equation}
A_{mm'}^{11}=\left(A_{mm'}^{1'1'}\right)^{*},\,A_{mm'}^{1'1'}=\left(A_{mm'}^{11}\right)^{*},\,A_{mm'}^{11'}=\left(A_{mm'}^{11'}\right)^{*},\,A_{mm'}^{1'1}=\left(A_{mm'}^{1'1}\right)^{*}.
\end{equation}
In other words, $A_{mm'}^{11'}$ and $A_{mm'}^{1'1}$ are real numbers while $A_{mm'}^{11}$ and $A_{mm'}^{1'1'}$ are complex conjugate.

We thus can easily write the real part $\rho_{mm'r}$ of $\rho_{mm'}$: 
\begin{equation}
\rho_{mm'r}=\frac{1}{2}\left[\left(A_{mm'}^{11}+A_{mm'}^{1'1'}\right)\left(\rho_{11}+\rho_{1'1'}\right)+\left(A_{mm'}^{11'}+A_{mm'}^{1'1}\right)\left(\rho_{11'}+\rho_{1'1}\right)\right].
\end{equation}
Since at equilibrium, $\rho_{mm'r}^{\left(eq\right)}=\rho_{11'}^{\left(eq\right)}=\rho_{1'1}^{\left(eq\right)}\approx0$ whereas $\rho_{11}^{\left(eq\right)}\approx\rho_{1'1'}^{\left(eq\right)}\ne0$, we have $\left(A_{mm'}^{11}+A_{mm'}^{1'1'}\right)\approx0$. Thus, 
\begin{equation}
\rho_{mm'r}=A_{mm'}^{11'r}\rho_{11'r},\label{eq:rhomm'r}
\end{equation}
where we have defined $A_{mm'}^{11'r}\equiv A_{mm'}^{11'}+A_{mm'}^{1'1}$.

Similarly for $\rho_{mm'i}$, we have 
\begin{align}
\rho_{mm'i} & =A_{mm'}^{11i}\left(\rho_{11}-\rho_{1'1'}\right)+A_{mm'}^{11'i}\rho_{11'i},\label{eq:form_of_rho_mm'i}
\end{align}
where $A_{mm'}^{11i}\equiv\mathrm{Im}\left(A_{mm'}^{11}\right)$ and $A_{mm}^{11'i}\equiv A_{mm'}^{11'}-A_{mm'}^{1'1}$.

\subsection{Summary}

In summary we have the following approximations for density matrix elements of the excited doublets/singlets in stationary limit: 
\begin{gather}
\rho_{mm}=A_{mm}^{11}\rho_{11}+A_{mm}^{1'1'}\rho_{1'1'}+A_{mm}^{11'i}\rho_{11'i},\\
\rho_{m'm'}=A_{mm}^{11}\rho_{1'1'}+A_{mm}^{1'1'}\rho_{11}-A_{mm}^{11'i}\rho_{11'i},\\
\rho_{mm'r}=A_{mm'}^{11'r}\rho_{11'r},\\
\rho_{mm'i}=A_{mm'}^{11i}\left(\rho_{11}-\rho_{1'1'}\right)+A_{mm'}^{11'i}\rho_{11'i},
\end{gather}
where all above $A_{mn}^{ij}$ are real parameters.

\section{Redfield master equation for the ground doublet}

From the semi-secular approximated Redfield equation for the general $d\rho_{mm}/dt$ and $d\rho_{mm'}/dt$ in the main text, the equation for density matrix elements of the ground doublet can be straightforwardly derived: 

\begin{align}
\frac{d\rho_{11}}{dt} & =\left(\Delta_{1i}\rho_{11'r}-\Delta_{1r}\rho_{11'i}\right)+\sum_{n^{\mathrm{th}}}\left(R_{11,nn}\rho_{nn}+R_{11,n'n'}\rho_{n'n'}\right)+2\sum_{n^{\mathrm{th}}}\left(R_{11,nn'}^{r}\rho_{nn'r}-R_{11,nn'}^{i}\rho_{nn'i}\right),\label{eq:drho11/dt}\\
\frac{d\rho_{1'1'}}{dt} & =-\left(\Delta_{1i}\rho_{11'r}-\Delta_{1r}\rho_{11'i}\right)+\sum_{n^{\mathrm{th}}}\left(R_{1'1',nn}\rho_{nn}+R_{1'1',n'n'}\rho_{n'n'}\right)+2\sum_{n^{\mathrm{th}}}\left(R_{1'1',nn'}^{r}\rho_{nn'r}-R_{1'1',nn'}^{i}\rho_{nn'i}\right),\\
\frac{d\rho_{11'r}}{dt} & =-\frac{\Delta_{1i}}{2}\left(\rho_{11}-\rho_{1'1'}\right)+W_{1}\rho_{11'i}+\sum_{n^{\mathrm{th}}}\left(R_{11',nn}^{r}\rho_{nn}+R_{11',n'n'}^{r}\rho_{n'n'}\right)\nonumber \\
 & \qquad\qquad\qquad\qquad\qquad\qquad\qquad\qquad+\sum_{n^{\mathrm{th}}}\left[\left(R_{11',nn'}^{r}+R_{1'1,nn'}^{r}\right)\rho_{nn'r}-\left(R_{11',nn'}^{i}+R_{1'1,nn'}^{i}\right)\rho_{nn'i}\right],\label{eq:drho11'r/dt-1}\\
\frac{d\rho_{11'i}}{dt} & =\frac{\Delta_{1r}}{2}\left(\rho_{11}-\rho_{1'1'}\right)-W_{1}\rho_{11'r}+\sum_{n^{\mathrm{th}}}\left(R_{11',nn}^{i}\rho_{nn}+R_{11',n'n'}^{i}\rho_{n'n'}\right)\nonumber \\
 & \qquad\qquad\qquad\qquad\qquad\qquad\qquad\qquad+\sum_{n^{\mathrm{th}}}\left[\left(R_{11',nn'}^{i}-R_{1'1,nn'}^{i}\right)\rho_{nn'r}+\left(R_{11',nn'}^{r}-R_{1'1,nn'}^{r}\right)\rho_{nn'i}\right].
\end{align}
where $R_{ij,mn}$ are matrix elements of the Redfield super-operator \citep{Garanin2011,Ho2017}.

\subsection{Diagonal terms}

Substituting $\rho_{mm}$, $\rho_{m'm'}$, $\rho_{mm'r}$, and $\rho_{mm'i}$ from the above section, Eqs. (\ref{eq:form_of_rho_mm}, \ref{eq:form_of_rho_m'm'}, \ref{eq:rhomm'r}, \ref{eq:form_of_rho_mm'i}) into the semi-secular approximated Redfield equation of $d\rho_{11}/dt$, Eq. \eqref{eq:drho11/dt} then simplifying gives:
\begin{align}
\frac{d\rho_{11}}{dt} & =\left(\Delta_{1i}\rho_{11'r}-\Delta_{1r}\rho_{11'i}\right)+\sum_{n^{\mathrm{th}}}\left(R_{11,nn}\rho_{nn}+R_{11,n'n'}\rho_{n'n'}\right)+2\sum_{n^{\mathrm{th}}}R_{11,nn'}\rho_{nn'r}\nonumber \\
 & =\left(\Delta_{1i}+2R_{11,11'}^{r}+2\sum_{n^{\mathrm{th}}\ne1^{\mathrm{st}}}R_{11,nn'}^{r}A_{nn'}^{11'r}\right)\rho_{11'r}\nonumber \\
 & \qquad+\left[-\Delta_{1r}+\sum_{n^{\mathrm{th}}\ne1^{\mathrm{st}}}\left(R_{11,nn}-R_{11,n'n'}\right)A_{nn}^{11'i}-2R_{11,11'}^{i}-2\sum_{n^{\mathrm{th}}\ne1^{\mathrm{st}}}R_{11,nn'}^{i}A_{nn'}^{11'i}\right]\rho_{11'i}\nonumber \\
 & \qquad+\left[R_{11,11}+\sum_{n^{\mathrm{th}}\ne1^{\mathrm{st}}}\left(R_{11,nn}A_{nn}^{11}+R_{11,n'n'}A_{nn}^{1'1'}\right)-2\sum_{n^{\mathrm{th}}\ne1^{\mathrm{st}}}R_{11,nn'}^{i}A_{nn'}^{11i}\right]\rho_{11}\nonumber \\
 & \qquad+\left[R_{11,1'1'}+\sum_{n^{\mathrm{th}}\ne1^{\mathrm{st}}}\left(R_{11,nn}A_{nn}^{1'1'}+R_{11,n'n'}A_{nn}^{11}\right)+2\sum_{n^{\mathrm{th}}\ne1^{\mathrm{st}}}R_{11,nn'}^{i}A_{nn'}^{11i}\right]\rho_{1'1'}\nonumber \\
 & \equiv\Delta_{1i}^{\prime}\rho_{11'r}-\Delta_{1r}^{\prime}\rho_{11'i}+\frac{1}{2}\left(\Gamma_{e}\rho_{1'1'}-\Gamma_{e}^{\prime}\rho_{11}\right).\label{eq:drho11-4vars}
\end{align}

Similarly we have for $d\rho_{1'1'}/dt$: 
\begin{align}
\frac{d\rho_{1'1'}}{dt} & =\left(-\Delta_{1i}+2R_{1'1',11'}^{r}+2\sum_{n^{\mathrm{th}}\ne1^{\mathrm{st}}}R_{1'1',nn'}^{r}A_{nn'}^{11'r}\right)\rho_{11'r}\nonumber \\
 & \qquad+\left[\Delta_{1r}-2R_{1'1',11'}^{i}+\sum_{n^{\mathrm{th}}\ne1^{\mathrm{st}}}\left(R_{1'1',nn}-R_{11,nn}\right)A_{nn}^{11'i}-2\sum_{n^{\mathrm{th}}\ne1^{\mathrm{st}}}R_{1'1',nn'}^{i}A_{nn'}^{11'i}\right]\rho_{11'i}\nonumber \\
 & \qquad+\left[R_{1'1',11}+\sum_{n^{\mathrm{th}}\ne1^{\mathrm{st}}}\left(R_{1'1',nn}A_{nn}^{11}+R_{1'1',n'n'}A_{nn}^{1'1'}\right)-2\sum_{n^{\mathrm{th}}\ne1^{\mathrm{st}}}R_{1'1',nn'}^{i}A_{nn'}^{11i}\right]\rho_{11}\nonumber \\
 & \qquad+\left[R_{1'1',1'1'}+\sum_{n^{\mathrm{th}}\ne1^{\mathrm{st}}}\left(R_{1'1',nn}A_{nn}^{1'1'}+R_{1'1',n'n'}A_{nn}^{11}\right)+2\sum_{n^{\mathrm{th}}\ne1^{\mathrm{st}}}R_{1'1',nn'}^{i}A_{nn'}^{11i}\right]\rho_{1'1'}\nonumber \\
 & \equiv-\Delta_{1i}^{\prime\prime}\rho_{11'r}+\Delta_{1r}^{\prime\prime}\rho_{11'i}+\frac{1}{2}\left(\Gamma_{e}^{\prime\prime}\rho_{11}-\Gamma_{e}^{\prime\prime\prime}\rho_{1'1'}\right).\label{eq:drho1'1'-4vars}
\end{align}

At equilibrium we have $d\rho_{11}/dt=d\rho_{1'1'}/dt=0$. Supposing that the energy bias of the ground doublet is much smaller than the temperature, i.e. $\rho_{11}^{\left(eq\right)}=\rho_{11}^{\left(eq\right)}$ and $\rho_{11'r}^{\left(eq\right)}=\rho_{11'i}^{\left(eq\right)}=0$, we further obtain:
\begin{gather}
\Gamma_{e}=\Gamma_{e}^{\prime},\\
\text{\ensuremath{\Gamma_{e}^{\prime\prime}}=\ensuremath{\Gamma_{e}^{\prime\prime\prime}}},
\end{gather}

In addition, within the stationary limit we have  $d\left(\rho_{11}+\rho_{1'1'}\right)/dt=0$ at any time $t$, it is straightforward from Eqs. (\ref{eq:drho11-4vars}-\ref{eq:drho1'1'-4vars}) that 
\begin{gather}
\Delta_{1r}^{\prime\prime}=\Delta_{1r}^{\prime},\\
\Delta_{1i}^{\prime\prime}=\Delta_{1i}^{\prime},\\
\Gamma_{e}=\Gamma_{e}^{\prime}=\Gamma_{e}^{\prime\prime}=\Gamma_{e}^{\prime\prime\prime}.
\end{gather}

Accordingly, we can reduce Eq. \eqref{eq:drho11-4vars} and \eqref{eq:drho1'1'-4vars} further to obtain:
\begin{align}
\frac{d\rho_{11}}{dt} & =\Delta_{1i}^{\prime}\rho_{11'r}-\Delta_{1r}^{\prime}\rho_{11'i}+\frac{1}{2}\Gamma_{e}\left(\rho_{1'1'}-\rho_{11}\right),\label{eq:drho11/dt-final}\\
\frac{d\rho_{1'1'}}{dt} & =-\Delta_{1i}^{\prime}\rho_{11'r}+\Delta_{1r}^{\prime}\rho_{11'i}+\frac{1}{2}\Gamma_{e}\left(\rho_{11}-\rho_{1'1'}\right).\label{eq:drho1'1'/dt-final}
\end{align}
where: 
\begin{align}
\Delta_{1r}^{\prime} & \equiv\Delta_{1r}-\sum_{n^{\mathrm{th}}\ne1^{\mathrm{st}}}\left(R_{11,nn}-R_{11,n'n'}\right)A_{nn}^{11'i}+2R_{11,11'}^{i}+2\sum_{n^{\mathrm{th}}\ne1^{\mathrm{st}}}R_{11,nn'}^{i}A_{nn'}^{11'i}\approx\Delta_{1r},\\
\Delta_{1i}^{\prime} & \equiv\Delta_{1i}+2R_{11,11'}^{r}+2\sum_{n^{\mathrm{th}}\ne1^{\mathrm{st}}}R_{11,nn'}^{r}A_{nn'}^{11'r}\approx\Delta_{1i},\\
\Gamma_{e} & =2\left[R_{11,1'1'}+\sum_{n^{\mathrm{th}}\ne1^{\mathrm{st}}}\left(R_{11,n'n'}A_{nn}^{11}+R_{11,nn}A_{nn}^{1'1'}\right)+2\sum_{n^{\mathrm{th}}\ne1^{\mathrm{st}}}R_{11,nn'}^{i}A_{nn'}^{11i}\right].\label{eq:Gamma_e}
\end{align}
and the approximated version: 
\begin{align}
\frac{d\rho_{11}}{dt} & =\Delta_{1i}\rho_{11'r}-\Delta_{1r}\rho_{11'i}+\frac{1}{2}\Gamma_{e}\left(\rho_{1'1'}-\rho_{11}\right),\label{eq:drho11/dt-final-2}\\
\frac{d\rho_{1'1'}}{dt} & =-\Delta_{1i}\rho_{11'r}+\Delta_{1r}\rho_{11'i}+\frac{1}{2}\Gamma_{e}\left(\rho_{11}-\rho_{1'1'}\right).\label{eq:drho1'1'/dt-final-2}
\end{align}

\subsection{Off-diagonal terms}

Similar to $d\rho_{11}/dt$ and $d\rho_{1'1'}/dt$, by substituting $\rho_{mm}$, $\rho_{m'm'}$, $\rho_{mm'r}$, and $\rho_{mm'i}$ from Eqs. (\ref{eq:form_of_rho_mm}, \ref{eq:form_of_rho_m'm'}, \ref{eq:rhomm'r}, \ref{eq:form_of_rho_mm'i}) into the semi-secular approximated Redfield equation of $d\rho_{11'r}/dt$, we obtain the simplified equation for $d\rho_{11'r}/dt$:

\begin{align}
\frac{d\rho_{11'r}}{dt} & =\left[-\frac{\Delta_{1i}}{2}+R_{11',11}^{r}+\sum_{n^{\mathrm{th}}\ne1^{\mathrm{st}}}\left(R_{11',nn}^{r}A_{nn}^{11}+R_{11',n'n'}^{r}A_{nn}^{1'1'}\right)-\left(R_{11',nn'}^{i}+R_{1'1,nn'}^{i}\right)A_{nn'}^{11i}\right]\rho_{11}\nonumber \\
 & \qquad+\left[\frac{\Delta_{1i}}{2}+R_{11',1'1'}^{r}+\sum_{n^{\mathrm{th}}\ne1^{\mathrm{st}}}\left(R_{11',nn}^{r}A_{nn}^{1'1'}+R_{11',n'n'}^{r}A_{nn}^{11}\right)+\left(R_{11',nn'}^{i}+R_{1'1,nn'}^{i}\right)A_{nn'}^{11i}\right]\rho_{1'1'}\nonumber \\
 & \qquad+\left[R_{11',11'}^{r}+R_{1'1,11'}^{r}+\sum_{n^{\mathrm{th}}\ne1^{\mathrm{st}}}\left(R_{11',nn'}^{r}+R_{1'1,nn'}^{r}\right)A_{nn'}^{11'r}\right]\rho_{11'r}\nonumber \\
 & \qquad+\left[W_{1}-\left(R_{11',11'}^{i}+R_{1'1,11'}^{i}\right)+\sum_{n^{\mathrm{th}}\ne1^{\mathrm{st}}}\left(R_{11',nn}^{r}A_{nn}^{11'i}-R_{11',n'n'}^{r}A_{nn}^{11'i}\right)-\left(R_{11',nn'}^{i}+R_{1'1,nn'}^{i}\right)A_{nn'}^{11'i}\right]\rho_{11'i}\nonumber \\
\equiv & -\frac{\Delta_{1i}^{\prime\prime\prime}}{2}\left(\rho_{11}-\rho_{1'1'}\right)-\gamma_{11'}^{\prime}\rho_{11'r}+W_{1}^{\prime}\rho_{11'i}.\label{eq:drho11'r/dt}
\end{align}
where
\begin{align}
\Delta_{1i}^{\prime\prime\prime} & \equiv\Delta_{1i}-2R_{11',11}^{r}+2\sum_{n^{\mathrm{th}}\ne1^{\mathrm{st}}}\left[\left(R_{11',nn'}^{i}+R_{1'1,nn'}^{i}\right)A_{nn'}^{11i}-\left(R_{11',nn}^{r}A_{nn}^{11}+R_{11',n'n'}^{r}A_{nn}^{1'1'}\right)\right]\approx\Delta_{1i}\\
\gamma_{11'}^{\prime} & \equiv-\left[R_{11',11'}^{r}+R_{1'1,11'}^{r}+\sum_{n^{\mathrm{th}}\ne1^{\mathrm{st}}}\left(R_{11',nn'}^{r}+R_{1'1,nn'}^{r}\right)A_{nn'}^{11'r}\right]\approx-R_{11',11'}^{r}\equiv\gamma_{11'}\\
W_{1}^{\prime} & \equiv W_{1}-\left(R_{11',11'}^{i}+R_{1'1,11'}^{i}\right)+\sum_{n^{\mathrm{th}}\ne1^{\mathrm{st}}}\left(R_{11',nn}^{r}A_{nn}^{11'i}-R_{11',n'n'}^{r}A_{nn}^{11'i}\right)-\left(R_{11',nn'}^{i}+R_{1'1,nn'}^{i}\right)A_{nn'}^{11'i}\approx W_{1}
\end{align}

Meanwhile, the equation for $d\rho_{11'i}/dt$ is: 
\begin{align}
\frac{d\rho_{11'i}}{dt} & =\left[\frac{\Delta_{1r}}{2}+R_{11',11}^{i}+\sum_{n^{\mathrm{th}}\ne1^{\mathrm{st}}}\left(R_{11',nn}^{i}A_{nn}^{11}+R_{11',n'n'}^{i}A_{nn}^{1'1'}\right)+\left(R_{11',nn'}^{r}-R_{1'1,nn'}^{r}\right)A_{nn'}^{11i}\right]\rho_{11}\nonumber \\
 & +\left[-\frac{\Delta_{1r}}{2}+R_{11',1'1'}^{i}+\sum_{n^{\mathrm{th}}\ne1^{\mathrm{st}}}\left(R_{11',nn}^{i}A_{nn}^{1'1'}+R_{11',n'n'}^{i}A_{nn}^{11}\right)-\left(R_{11',nn'}^{r}-R_{1'1,nn'}^{r}\right)A_{nn'}^{11i}\right]\rho_{1'1'}\nonumber \\
 & +\left[-W_{1}+\left(R_{11',11'}^{i}-R_{1'1,11'}^{i}\right)+\sum_{n^{\mathrm{th}}\ne1^{\mathrm{st}}}\left(R_{11',nn'}^{i}-R_{1'1,nn'}^{i}\right)A_{nn'}^{11'r}\right]\rho_{11'r}\nonumber \\
 & +\left[\left(R_{11',11'}^{r}-R_{1'1,11'}^{r}\right)+\sum_{n^{\mathrm{th}}\ne1^{\mathrm{st}}}\left(R_{11',nn}^{i}A_{nn}^{11'i}-R_{11',n'n'}^{i}A_{nn}^{11'i}\right)+\left(R_{11',nn'}^{r}-R_{1'1,nn'}^{r}\right)A_{nn'}^{11'i}\right]\rho_{11'i}\nonumber \\
 & \equiv\frac{\Delta_{1r}^{\prime\prime\prime}}{2}\left(\rho_{11}-\rho_{1'1'}\right)-\gamma_{11'}^{\prime\prime}\rho_{11'i}-W_{1}^{\prime\prime}\rho_{11'r}.\label{eq:drho11'i/dt}
\end{align}
where 
\begin{align}
\Delta_{1r}^{\prime\prime\prime} & \equiv\Delta_{1r}+2R_{11',11}^{i}+2\sum_{n^{\mathrm{th}}\ne1^{\mathrm{st}}}\left[\left(R_{11',nn}^{i}A_{nn}^{11}+R_{11',n'n'}^{i}A_{nn}^{1'1'}\right)+\left(R_{11',nn'}^{r}-R_{1'1,nn'}^{r}\right)A_{nn'}^{11i}\right]\approx\Delta_{1r}\\
\gamma_{11'}^{\prime\prime} & \equiv-\left[\left(R_{11',11'}^{r}-R_{1'1,11'}^{r}\right)+\sum_{n^{\mathrm{th}}\ne1^{\mathrm{st}}}\left(R_{11',nn}^{i}A_{nn}^{11'i}-R_{11',n'n'}^{i}A_{nn}^{11'i}\right)+\left(R_{11',nn'}^{r}-R_{1'1,nn'}^{r}\right)A_{nn'}^{11'i}\right]\approx-R_{11',11'}^{r}\equiv\gamma_{11'}\\
W_{1}^{\prime\prime} & \equiv W_{1}-\left(R_{11',11'}^{i}-R_{1'1,11'}^{i}\right)-\sum_{n^{\mathrm{th}}\ne1^{\mathrm{st}}}\left(R_{11',nn'}^{i}-R_{1'1,nn'}^{i}\right)A_{nn'}^{11'r}\approx W_{1}
\end{align}

The approximated equations of $d\rho_{11'r}/dt$ and $d\rho_{11'i}/dt$ then are: 
\begin{align}
\frac{d\rho_{11'r}}{dt} & =-\frac{\Delta_{1i}}{2}\left(\rho_{11}-\rho_{1'1'}\right)-\gamma_{11'}\rho_{11'r}+W_{1}\rho_{11'i},\\
\frac{d\rho_{11'i}}{dt} & =\frac{\Delta_{1r}}{2}\left(\rho_{11}-\rho_{1'1'}\right)-\gamma_{11'}\rho_{11'i}-W_{1}\rho_{11'r}.
\end{align}

\subsection{Summary}

Within the stationary limit for excited doublets/singlets, the approximated Redfield master equation for the ground doublet, and relaxation of the spin system, can be summarized as follows: 
\begin{align}
\frac{d\rho_{11}}{dt} & =\Delta_{1i}\rho_{11'r}-\Delta_{1r}\rho_{11'i}+\frac{1}{2}\Gamma_{e}\left(\rho_{1'1'}-\rho_{11}\right),\label{eq:drho11/dt-final-1}\\
\frac{d\rho_{1'1'}}{dt} & =-\Delta_{1i}\rho_{11'r}+\Delta_{1r}\rho_{11'i}+\frac{1}{2}\Gamma_{e}\left(\rho_{11}-\rho_{1'1'}\right),\label{eq:drho1'1'/dt-final-1}\\
\frac{d\rho_{11'r}}{dt} & =-\frac{\Delta_{1i}}{2}\left(\rho_{11}-\rho_{1'1'}\right)-\gamma_{11'}\rho_{11'r}+W_{1}\rho_{11'i},\\
\frac{d\rho_{11'i}}{dt} & =\frac{\Delta_{1r}}{2}\left(\rho_{11}-\rho_{1'1'}\right)-\gamma_{11'}\rho_{11'i}-W_{1}\rho_{11'r}.
\end{align}
where 
\begin{align*}
\Gamma_{e} & =2\left[R_{11,1'1'}+\sum_{n^{\mathrm{th}}\ne1^{\mathrm{st}}}\left(R_{11,n'n'}A_{nn}^{11}+R_{11,nn}A_{nn}^{1'1'}\right)+2\sum_{n^{\mathrm{th}}\ne1^{\mathrm{st}}}R_{11,nn'}^{i}A_{nn'}^{11i}\right].
\end{align*}

As the states $\ket{1}$ and $\ket{1'}$ have opposite magnetic moments, in order to find the magnetization relaxation rate of the spin system, we can change the variables $\rho_{11}$ and $\rho_{1'1'}$ to $X_{1}\equiv\rho_{11}-\rho_{1'1'}$. The above differential equations can thus be rewritten as: 
\begin{gather}
\frac{dX_{1}}{dt}=-\Gamma_{e}X_{1}-2\left(\Delta_{1r}\rho_{11'i}-\Delta_{1i}\rho_{11'r}\right),\label{eq:dX1}\\
\frac{d\rho_{11'r}}{dt}=-\gamma_{11'}\rho_{11'r}+W_{1}\rho_{11'i}-\frac{\Delta_{1i}}{2}X_{1},\label{eq:drho11'r}\\
\frac{d\rho_{11'i}}{dt}=-W_{1}\rho_{11'r}-\gamma_{11'}\rho_{11'i}+\frac{\Delta_{1r}}{2}X_{1},\label{eq:drho11'i}
\end{gather}
or in the matrix form:

\begin{equation}
\frac{d\vt{\rho}}{dt}=\Phi\vt{\rho},
\end{equation}
where the transition rate matrix $\Phi$ and $\vt{\rho}$, respectively, are: 
\begin{equation}
\Phi=\left(\begin{array}{ccc}
-\Gamma_{e} & 2\Delta_{1i} & -2\Delta_{1r}\\
-\Delta_{1i}/2 & -\gamma_{11'} & W_{1}\\
\Delta_{1r}/2 & -W_{1} & -\gamma_{11'}
\end{array}\right),\,\vt{\rho}=\left(\begin{array}{c}
X_{1}\\
\rho_{11'r}\\
\rho_{11'i}
\end{array}\right).\label{eq:transition_rate_matrix}
\end{equation}
Three eigenvalues of the matrix $\Phi$ are corresponding rates of three relaxation modes.

We also have the corresponding characteristic polynomial: 
\begin{equation}
\lambda^{3}-\lambda^{2}(2\gamma_{11'}+\Gamma_{e})+\lambda\left(\gamma_{11'}^{2}+2\gamma_{11'}\Gamma_{e}+\Omega_{1}^{2}\right)-\left(\gamma_{11'}^{2}\Gamma_{e}+\gamma_{11'}\Delta_{1}^{2}+\Gamma_{e}W_{1}^{2}\right)=0,\label{eq:characteristic poly}
\end{equation}
where 
\begin{align}
\Omega_{1} & =\sqrt{\Delta_{1}^{2}+W_{1}^{2}},\\
\Delta_{1}^{2} & =\Delta_{1r}^{2}+\Delta_{1i}^{2}.
\end{align}

\section{Magnetization relaxation of a molecular spin at resonance\label{sec:Magnetization-relaxation-at-resonance}}

Relaxation of the magnetization at resonance $W_{1}=0$ can be obtained by diagonalizing the transition rate matrix Eq. \eqref{eq:transition_rate_matrix}. Defining $\gamma\equiv\left(\gamma_{11'}^{\prime}-\Gamma_{e}\right)/2$, the corresponding relaxation rates and eigenvectors then are: 
\begin{alignat}{2}
\lambda_{1} & =\Gamma_{e}+\gamma-\sqrt{\gamma^{2}-\Delta_{1}^{2}},\qquad & \qquad\vt{\rho}_{1} & =\left(2\frac{\gamma+\sqrt{\gamma^{2}-\Delta_{1}^{2}}}{\Delta_{1r}},-\frac{\Delta_{1i}}{\Delta_{1r}},1\right),\\
\lambda_{2} & =\Gamma_{e}+\gamma+\sqrt{\gamma^{2}-\Delta_{1}^{2}},\qquad\qquad & \vt{\rho}_{2} & =\left(2\frac{\gamma-\sqrt{\gamma^{2}-\Delta_{1}^{2}}}{\Delta_{1r}},-\frac{\Delta_{1i}}{\Delta_{1r}},1\right),\\
\lambda_{3} & =\gamma_{11'}^{\prime},\qquad\qquad & \vt{\rho}_{3} & =\left(0,\frac{\Delta_{1r}}{\Delta_{1i}},1\right).
\end{alignat}

If at $t=0$, the system is prepared so that $X_{1}=1$ and $\rho_{11'r}=\rho_{11'i}=0$, we have 
\begin{gather}
\left(\begin{array}{c}
X_{1}\\
\rho_{11'r}\\
\rho_{11'i}
\end{array}\right)=\frac{\Delta_{1r}}{4\sqrt{\gamma^{2}-\Delta_{1}^{2}}}\left(\begin{array}{c}
2\frac{\gamma+\sqrt{\gamma^{2}-\Delta_{1}^{2}}}{\Delta_{1r}}e^{-\lambda_{1}t}-2\frac{\gamma-\sqrt{\gamma^{2}-\Delta_{1}^{2}}}{\Delta_{1r}}e^{-\lambda_{2}t}\\
\frac{\Delta_{1i}}{\Delta_{1r}}\left(e^{-\lambda_{2}t}-e^{-\lambda_{1}t}\right)\\
e^{-\lambda_{1}t}-e^{-\lambda_{2}t}
\end{array}\right).
\end{gather}

Accordingly, the magnetization as function of time is: 
\begin{align}
M\left(t\right) & \sim\mathrm{Tr}\left(S_{z}\rho\right)=\sum_{m^{\mathrm{th}}}s_{m}\left(\rho_{mm}-\rho_{m'm'}\right)\nonumber \\
 & =\left[s_{1}+\sum_{m^{\mathrm{th}}\ne1^{\mathrm{st}}}s_{m}\left(A_{mm}^{11}-A_{mm}^{1'1'}\right)\right]X_{1}+\left[2\sum_{m^{\mathrm{th}}\ne1^{\mathrm{st}}}s_{m}A_{mm}^{11'i}\right]\rho_{11'i}\nonumber \\
 & =\frac{1}{2\sqrt{\gamma^{2}-\Delta_{1}^{2}}}\left\{ \left(\gamma+\sqrt{\gamma^{2}-\Delta_{1}^{2}}\right)\left[s_{1}+\sum_{m^{\mathrm{th}}\ne1^{\mathrm{st}}}s_{m}\left(A_{mm}^{11}-A_{mm}^{1'1'}\right)\right]+\Delta_{1r}\sum_{m^{\mathrm{th}}\ne1^{\mathrm{st}}}s_{m}A_{mm}^{11'i}\right\} e^{-\lambda_{1}t}\nonumber \\
 & -\frac{1}{2\sqrt{\gamma^{2}-\Delta_{1}^{2}}}\left\{ \left(\gamma-\sqrt{\gamma^{2}-\Delta_{1}^{2}}\right)\left[s_{1}+\sum_{m^{\mathrm{th}}\ne1^{\mathrm{st}}}s_{m}\left(A_{mm}^{11}-A_{mm}^{1'1'}\right)\right]+\Delta_{1r}\sum_{m^{\mathrm{th}}\ne1^{\mathrm{st}}}s_{m}A_{mm}^{11'i}\right\} e^{-\lambda_{2}t}
\end{align}

Defining: 
\begin{align}
r & \equiv\sum_{m^{\mathrm{th}}\ne1^{\mathrm{st}}}\frac{s_{m}A_{mm}^{11'i}}{s_{1}+\sum_{m^{\mathrm{th}}\ne1^{\mathrm{st}}}s_{m}\left(A_{mm}^{11}-A_{mm}^{1'1'}\right)},\\
\Gamma_{1,2}^{\mathrm{tn}} & \equiv\gamma\mp\sqrt{\gamma^{2}-\Delta_{1}^{2}},
\end{align}
we obtain: 
\begin{align}
M\left(t\right) & =\frac{M_{0}e^{-\Gamma_{e}t}}{2\sqrt{\gamma^{2}-\Delta_{1}^{2}}}\left[\left(\Gamma_{2}^{\mathrm{tn}}+\Delta_{1r}r\right)e^{-\Gamma_{1}^{\mathrm{tn}}t}-\left(\Gamma_{1}^{\mathrm{tn}}+\Delta_{1r}r\right)e^{-\Gamma_{2}^{\mathrm{tn}}t}\right],\label{eq:M(t)-2}
\end{align}
where $M_{0}\equiv M\left(t=0\right)$. Considering that the correlation between the diagonal matrix elements of the excited states and the off-diagonal density matrix elements of the ground doublet $A_{mm}^{11'i}$ is typically weak, the above can be further approximated as: 
\begin{align}
M\left(t\right) & =\frac{M_{0}e^{-\Gamma_{e}t}}{2\sqrt{\gamma^{2}-\Delta_{1}^{2}}}\left[\Gamma_{2}^{\mathrm{tn}}e^{-\Gamma_{1}^{\mathrm{tn}}t}-\Gamma_{1}^{\mathrm{tn}}e^{-\Gamma_{2}^{\mathrm{tn}}t}\right].\label{eq:M(t)}
\end{align}

\subsection{Relaxation at transition point $\gamma=\Delta_{1}$}

At transition point $\gamma=\Delta_{1}$ (at some $T_{0}$), with the initial condition $X_{1}=1$ and $\rho_{11'r}=\rho_{11'i}=0$, we have 
\begin{gather}
X_{1}\left(t\right)=(1+\Delta_{1}t)e^{-\left(\Delta_{1}+\Gamma_{e}\right)t},\\
\rho_{11'r}=-\frac{1}{2}\Delta_{1i}te^{-\left(\Delta_{1}+\Gamma_{e}\right)t},\\
\rho_{11'i}=\frac{1}{2}\Delta_{1r}te^{-\left(\Delta_{1}+\Gamma_{e}\right)t},\\
M\left(t,T_{0}\right)=M_{0}e^{-\left(\Gamma_{e,0}+\Delta_{1}\right)t}\left[1+\Delta_{1}t\right].\label{eq:M(t,T_0)}
\end{gather}

\subsection{Relaxation in the positive proximity of the transition point $\gamma=\Delta_{1}+\delta\gamma$}

Supposing that $\gamma_{11'}\left(T_{0}+\delta T\right)=\gamma_{11',0}+\chi\delta T$, $\Gamma_{e}\left(T_{0}+\delta T\right)=\Gamma_{e,0}+\xi\delta T$, accordingly $\gamma=\frac{1}{2}\left[\left(\gamma_{11',0}-\Gamma_{e,0}\right)+\left(\chi-\xi\right)\delta T\right]$=$\Delta_{1}+\zeta\delta T$. From Eq. \eqref{eq:M(t)}), we have 
\begin{align}
M\left(t,T_{0}+\delta T\right) & =M\left(t,T_{0}\right)+\delta M\left(t,T_{0},\delta T\right)
\end{align}
where $M\left(t,T_{0}\right)$ from Eq. \eqref{eq:M(t,T_0)} and 
\begin{equation}
\delta M\left(t,T_{0},\delta T\right)\equiv M_{0}e^{-\left(\Gamma_{e,0}+\Delta_{1}\right)t}\delta Tt\left[\frac{1}{6}\left(\chi-\xi\right)\Delta_{1}^{2}t^{2}-\xi\Delta_{1}t-\xi\right],
\end{equation}
which allow us to find the time $\tau_{M}$ from which $\delta M\left(t\right)>0$: 
\begin{gather}
\tau_{M}=\frac{\sqrt{3}\xi}{\Delta_{1}\left(\chi-\xi\right)}\left(\sqrt{3}+\sqrt{1+\frac{2\chi}{\xi}}\right),
\end{gather}
The condition for $\tau_{M}$ in the observable time $\Delta_{1}^{-1}$ can be easily found 
\begin{equation}
\frac{\sqrt{3}\xi}{\left(\chi-\xi\right)}\left(\sqrt{3}+\sqrt{1+\frac{2\chi}{\xi}}\right)\le1\Longleftrightarrow\frac{\xi}{\chi}\le\frac{1}{13},
\end{equation}
or, 
\begin{equation}
\zeta/\xi\ge6.
\end{equation}

\section{Magnetization relaxation of a molecular spin in general case}

The quantity $X_{1}\left(t\right)$ in Eqs. (\ref{eq:dX1}-\ref{eq:drho11'i}) can be found in the form $X_{1}\left(t\right)=e^{-\Gamma_{e}t}\sum_{i=1}^{3}c_{i}e^{-\Gamma_{i}t}$ where $c_{i}$ depends on the initial conditions and $\Gamma_{i}$ are as follows \citep{Ho2022c}: 
\begin{align}
\Gamma_{1} & =\frac{4\gamma}{3}-\frac{3\Omega_{1}^{2}-4\gamma^{2}}{3S}+\frac{1}{3}S,\label{eq:Gamma_1}\\
\Gamma_{2} & =\frac{4\gamma}{3}+\frac{1+i\sqrt{3}}{6}\frac{3\Omega_{1}^{2}-4\gamma^{2}}{S}-\frac{1-i\sqrt{3}}{6}S,\\
\Gamma_{3} & =\frac{4\gamma}{3}+\frac{1-i\sqrt{3}}{6}\frac{3\Omega_{1}^{2}-4\gamma_{1}^{2}}{S}-\frac{1+i\sqrt{3}}{6}S,\label{eq:Gamma_3}
\end{align}
where 
\begin{align}
S\equiv & \sqrt[3]{9\gamma\left(\Delta_{1}^{2}-2W_{1}^{2}\right)-8\gamma^{3}+3\sqrt{3}\sqrt{D}},\\
D\equiv & 16\gamma^{4}W_{1}^{2}+\gamma^{2}\left(8W_{1}^{4}-20W_{1}^{2}\Delta_{1}^{2}-\Delta_{1}^{4}\right)+\Omega_{1}^{6}
\end{align}

Using the initial condition $\left(X_{1},\rho_{11'r},\rho_{11'i}\right)|_{t=0}=\left(1,0,0\right)$ we have the following equations for the constants $c_{i}$: 
\begin{gather}
\sum_{i=1}^{3}c_{i}=1,\quad\sum_{i=1}^{3}\Gamma_{i}c_{i}=0,\quad\sum\Gamma_{i}^{2}c_{i}=-\Delta_{1}^{2},
\end{gather}
which leads to \citep{Ho2022c}: 
\begin{align}
c_{1} & =\frac{\Gamma_{2}\Gamma_{3}-\Delta_{1}^{2}}{\left(\Gamma_{1}-\Gamma_{2}\right)\left(\Gamma_{1}-\Gamma_{3}\right)},\\
c_{2} & =\frac{\Gamma_{1}\Gamma_{3}-\Delta_{1}^{2}}{\left(\Gamma_{2}-\Gamma_{1}\right)\left(\Gamma_{2}-\Gamma_{3}\right)},\\
c_{3} & =\frac{\Gamma_{1}\Gamma_{2}-\Delta_{1}^{2}}{\left(\Gamma_{3}-\Gamma_{1}\right)\left(\Gamma_{3}-\Gamma_{2}\right)},
\end{align}
 where $\Gamma_{i}$, $i=1,2,3$, are given above. Time-dependent magnetization $M\left(t\right)$ of a molecular spin is approximately proportional to population difference in the ground doublet $X_{1}\left(t\right)$: 
\begin{align}
M\left(t\right) & \approx M_{0}e^{-\Gamma_{e}t}\left[\frac{\Gamma_{2}\Gamma_{3}-\Delta_{1}^{2}}{\left(\Gamma_{1}-\Gamma_{2}\right)\left(\Gamma_{1}-\Gamma_{3}\right)}e^{-\Gamma_{1}t}+\frac{\Gamma_{1}\Gamma_{3}-\Delta_{1}^{2}}{\left(\Gamma_{2}-\Gamma_{1}\right)\left(\Gamma_{2}-\Gamma_{3}\right)}e^{-\Gamma_{2}t}+\frac{\Gamma_{1}\Gamma_{2}-\Delta_{1}^{2}}{\left(\Gamma_{3}-\Gamma_{1}\right)\left(\Gamma_{3}-\Gamma_{2}\right)}e^{-\Gamma_{3}t}\right],\label{eq:M(t)-1}
\end{align}
where $\Gamma_{e}$ is given in Eq. \eqref{eq:Gamma_e} and $\Gamma_{i}$, $i=1,2,3,$ are given in Eqs. (\ref{eq:Gamma_1}-\ref{eq:Gamma_3}). Here, the first part $M_{0}e^{-\Gamma_{e}t}$ is magnetization relaxation without the tunneling effect in the ground doublet while the latter embraces this effect.

\section{Magnetization relaxation fitting: single-, bi-, or tri-exponential form?}

In general, magnetization relaxation in intermediate and low temperature regimes are a combination of three relaxation modes whose corresponding rates are a simple addition of  ``classical'' relaxation processes (Orbach, Raman, and direct process) $\Gamma_{e}$ and the ones resulting from the quantum tunneling in the ground (quasi) doublet. However, depending on the temperature and the energy bias in the ground doublet, one relaxation mode may still relax much slower than the remaining. Consequently, using one relaxation rate characterizing for the whole relaxation of the molecular spin may still be a good approximation for these cases. In contrast, from the resonance case $W_{1}=0$, it is clearly that in the proximity of the transition decoherence rate where $\gamma=\Delta_{1}$, relaxation of the magnetization must be described using a bi-exponential formula. Hence, the question is in which conditions the single-, bi-, or tri-exponential formula should be used. 

From our companion work \citep{Ho2022c}, we know that at the transition point $\gamma_{0}$ at least two lowest quantum tunneling rates share the same value of the real part. Just for a summary, here is dependence of the transition decoherence rate $\gamma_{0}$ on $\Delta_{1}$ and $W_{1}$: 
\begin{eqnarray}
\gamma_{0} & = & \begin{cases}
\Delta_{1} & \text{for\,\,\,}W_{1}=0,\\
\frac{\left(\sqrt{\Delta_{1}^{2}-8W_{1}^{2}}+3\Delta_{1}\right)\sqrt{4W_{1}^{2}-\Delta_{1}\sqrt{\Delta_{1}^{2}-8W_{1}^{2}}+\Delta_{1}^{2}}}{8\sqrt{2}W_{1}} & \text{for\,\,\,}0\le W_{1}<\frac{\Delta_{1}}{2\sqrt{2}},\\
\frac{3}{2\sqrt{2}}\sqrt{\Delta_{1}^{2}-2W_{1}^{2}} & \text{for\,\,\,}\frac{\Delta_{1}}{2\sqrt{2}}\le W_{1}<\frac{\Delta_{1}}{\sqrt{2}},\\
0 & \text{for\,\,\,}W_{1}\ge\frac{\Delta_{1}}{\sqrt{2}}.
\end{cases}\label{eq:gamma0}
\end{eqnarray}

\begin{figure}
\centering{}\includegraphics[width=10cm]{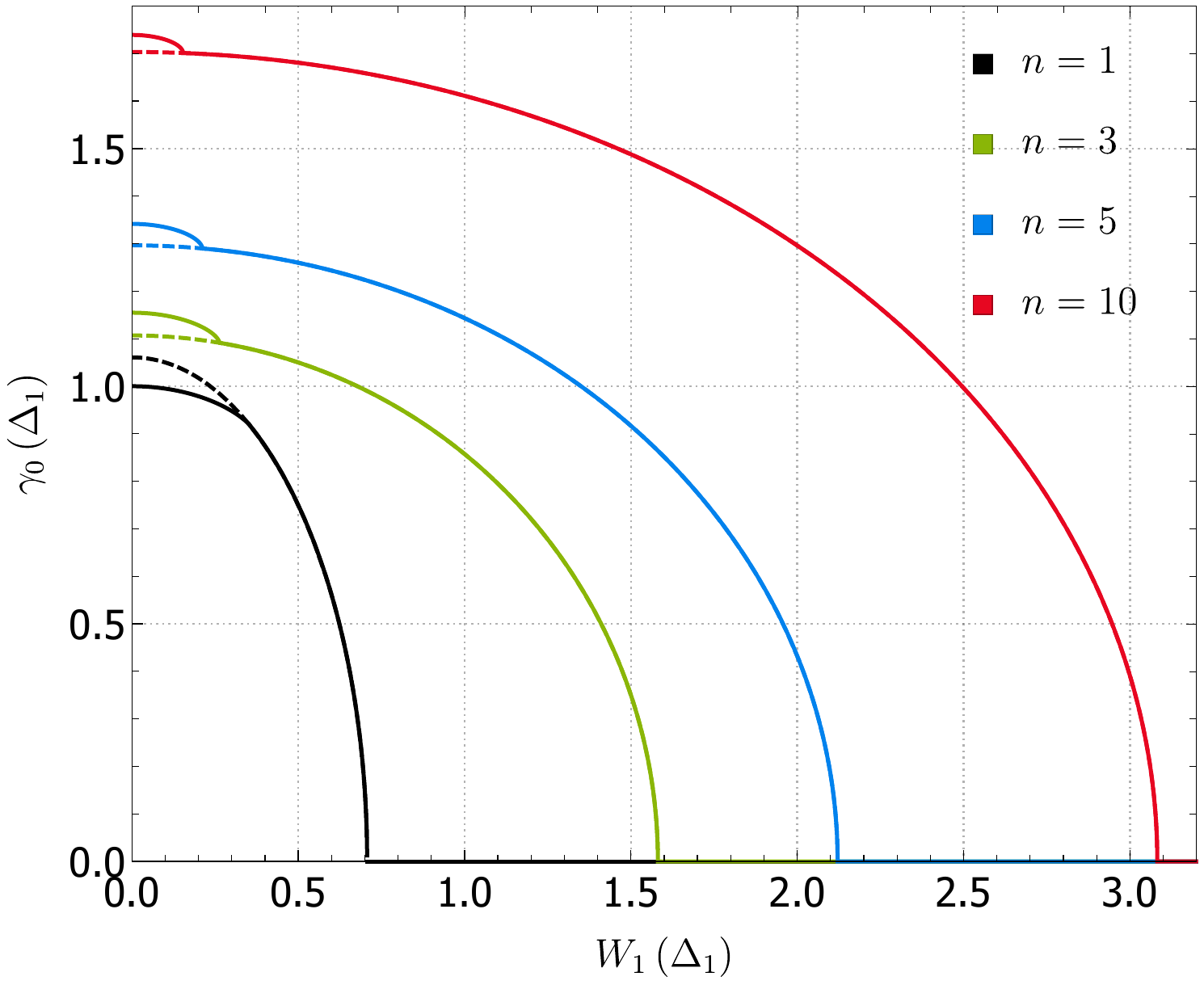}$\qquad\qquad$\caption{Transition decoherence rate $\gamma_{0}$ ($n=1$) and $\gamma_{n}$, $n=3,5,10$, as a function of the energy bias $W_{1}$ in the ground doublet in $\Delta_{1}=1$ unit. The dashed lines are $\gamma_{n}$ from Eq. \eqref{eq:gamma_n_point} applied for any $W_{1}$.\label{fig:Transition point}}
\end{figure}

As can be seen, for $W_{1}\le1/\sqrt{2}$, we can easily say that in the proximity of this special point $\gamma_{0}$, the magnetization relaxation should be fit using bi- or tri-exponential formula. However, remember that as long as (the real part of) relaxation rates are at the same order of magnitude then tri-exponential formula of the magnetization relaxation still needs to be used, we define $\gamma_{n}$ at which $\mathrm{Re}\left(\Gamma_{2,3}\right)=n\mathrm{Re}\left(\Gamma_{1}\right)\sim\mathcal{O}\left(\Gamma_{1}\right)$ where $1<n\le10$. From the governing equation of the magnetization relaxation in spin system above and/or its characteristic polynomial Eq. \eqref{eq:characteristic poly}, we can obtain the value for $\gamma_{n}$: 
\begin{equation}
\gamma_{n}=\frac{1}{2\sqrt{2}}\frac{2n+1}{2n-1}\mathrm{Re}\left[\sqrt{\left(2n-1\right)\Delta_{1}^{2}-2W_{1}^{2}}\right],\text{ for\,\,\,}W_{1}\ge\frac{\Delta_{1}}{2\sqrt{2}},\label{eq:gamma_n_point}
\end{equation}
from this value of $\gamma_{n}$, which corresponds to a specific temperature $T_{n}$, lowering temperature requires a tri-exponential formula for the description of the magnetization relaxation of a molecular spin. Unfortunately, for $W_{1}\le\Delta_{1}/2\sqrt{2}$, we do not have a simple formula for $\gamma_{n}$. We thus resort to a numerical visualization of $\gamma_{n}$. Fig. \ref{fig:Transition point} hence presents the dependence of $\gamma_{n}$ on $W_{1}$ in $\Delta_{1}$ unit for different value of $n$. 

As can be seen, the numerical calculation shows that Eq. \eqref{eq:gamma_n_point} (dashed lines) still gives a very good approximation of $\gamma_{n}$ at low value of $W_{1}$. Hence, we can take: 
\begin{align}
\gamma_{n} & \approx\frac{1}{2\sqrt{2}}\frac{2n+1}{2n-1}\mathrm{Re}\left[\sqrt{\left(2n-1\right)\Delta_{1}^{2}-2W_{1}^{2}}\right]\,\,\,\forall W_{1}.\label{eq:gamma_n_point2}
\end{align}
In summary, for the purpose of magnetization relaxation fitting:
\begin{itemize}
\item For $W_{1}=0$, the bi-exponential formula should be use in the proximity of the transition decoherence rate $\gamma_{0}=\Delta_{1}$ and for all $\gamma\le\gamma_{0}$.
\item For $0<W_{1}\le\Delta_{1}/\sqrt{2}$, the tri-exponential formula for the magnetization relaxation, Eq. \eqref{eq:M(t)-1}, needs to be used in the positive proximity of the transition decoherence rate $\gamma_{0}$ (from $\gamma_{0}$ to $\gamma_{n}$) and for all $\gamma\le\gamma_{0}$.
\item For $W_{1}>\Delta_{1}/\sqrt{2}$, the tri-exponential formula for the magnetization relaxation needs to be used for $\gamma<\gamma_{n}$ where $\gamma_{n}$ is given by Eq. \eqref{eq:gamma_n_point}.
\item For all other cases, single-exponential formula can be used. 
\end{itemize}
It should be noted that when $W_{1}>\Delta_{1}/2\sqrt{2}$, the governing equation for the quantum tunneling process gives two complex conjugates and one real root, experimental data can be fitted either by using Eq. \eqref{eq:M(t)-1} or another equivalent one, but easier for numerical fitting, whose all parameters are real numbers:
\begin{multline}
M\left(t\right)=M_{0}e^{-\Gamma_{e}t}\left[\frac{\Gamma_{2r}^{2}+\Gamma_{2i}^{2}-\Delta_{1}^{2}}{\left(\Gamma_{1}-\Gamma_{2r}\right)^{2}+\Gamma_{2i}^{2}}e^{-\Gamma_{1}t}+\left(1-\frac{\left(\Gamma_{2i}^{2}+\Gamma_{2r}^{2}\right)-\Delta_{1}^{2}}{\left(\Gamma_{1}-\Gamma_{2r}\right)^{2}+\Gamma_{2i}^{2}}\right)e^{-\Gamma_{2r}t}\cos\Gamma_{2i}t\right.\\
\left.+\frac{\Gamma_{1}\left(\Gamma_{1}\Gamma_{2r}-\Gamma_{2r}^{2}+\Gamma_{2i}^{2}\right)-\left(\Gamma_{1}-\Gamma_{2r}\right)\Delta_{1}^{2}}{\Gamma_{2i}\left[\left(\Gamma_{1}-\Gamma_{2r}\right)^{2}+\Gamma_{2i}^{2}\right]}e^{-\Gamma_{2r}t}\sin\Gamma_{2i}t\right]\\
\equiv M_{0}\left[(1-a)e^{-\left(\Gamma_{1}+\Gamma_{e}\right)t}+e^{-\left(\Gamma_{2r}+\Gamma_{e}\right)t}\left(a\cos\Gamma_{2i}t+b\sin\Gamma_{2i}t\right)\right].
\end{multline}
In principle, from the fitted parameter, the tunneling splitting of the ground doublet can be found. 

\bibliographystyle{apsrev4-1}
\bibliography{references}